\documentclass{amsart} 
\usepackage{graphicx}
\usepackage{amssymb, amsmath, sidecap}
\usepackage{amsfonts}
\usepackage{amssymb}
\usepackage{float}
\usepackage[version=3]{mhchem}
\usepackage{endnotes}

\newcommand{\R}{\mathbb R}

\def\la{\label}

\def\qtf{\qquad \text{ for }}
\def\cup{\bigcup}
\def\cap{\bigcap}  	
\def\R{\mathbb R}

\def\ei{\end{itemize}}
\def\bi{\begin{itemize}}
\def\bt{\begin{thm}}
\def\et{\end{thm}}
\def\bl{\begin{lem}}
\def\el{\end{lem}}
\def\bd{\begin{defi}}
\def\ed{\end{defi}}
\def\bc{\begin{cor}}
\def\ec{\end{cor}}
\def\bp{\begin{proof}}
\def\ep{\end{proof}}
\def\br{\begin{rem}}
\def\er{\end{rem}}

\def\bpp{\begin{proof}}
\def\epp{\end{proof}}

\def\bcon{\begin{conclusion}}
\def\econ{\end{conclusion}}

\newtheorem{thm}{Theorem}[section]
\newtheorem{lem}{Lemma}[section]
\newtheorem{defi}{Definition}[section]

\newtheorem{rem}{Remark}[section]
\newtheorem{cor}{Corollary}[section]

\newtheorem{conclusion}{Physical Conclusion}[section]

\numberwithin{equation}{section}
\numberwithin{figure}{section}

\begin{document}
\title{Dynamic  Transition and Pattern Formation in Taylor Problem}
\author[Ma]{Tian Ma}
\address[TM]{Department of Mathematics, Sichuan University,
Chengdu, P. R. China}

\author[Wang]{Shouhong Wang}
\address[SW]{Department of Mathematics,
Indiana University, Bloomington, IN 47405}
\email{showang@indiana.edu, http://www.indiana.edu/~fluid}

\thanks{The work was supported in part by grants from the
Office of Naval Research, the US National Science Foundation, and the Chinese National Science Foundation.}


\begin{abstract}
The main objective of this article is to study both dynamic and structural transitions of the Taylor-Couette flow, using the dynamic transition theory and geometric theory of incompressible flows developed recently by the authors. In particular we show that as the Taylor number crosses the critical number, the system undergoes either a continuous or a jump dynamic transition, dictated by the sign of a 
computable, nondimensional parameter $R$. In addition, we show that the new transition states have the Taylor vortex type of flow structure, which is structurally stable. 
\end{abstract}
\maketitle

\section{Introduction}
The study of hydrodynamic instability caused by the centrifugal
forces originated from the famous experiments conducted by \cite{taylor} in 1923, 
in which he observed and studied the
stability of an incompressible viscous fluid between two rotating
coaxial cylinders. In his experiments, Taylor investigated the case
where the gap between the two cylinders is small in comparison with
the mean radius, and the two cylinders rotate in the same direction.
He found that when the Taylor number $T$ is smaller than a critical
value $T_c>0$, called the critical Taylor number, the basic flow,
called the Couette flow, is stable, and when the Taylor number
crosses the critical value, the Couette flow breaks out into a radially symmetric
cellular pattern as in Figure \ref{f9.8}.
\begin{SCfigure}[25][t]
  \centering
  \includegraphics[width=0.35\textwidth]{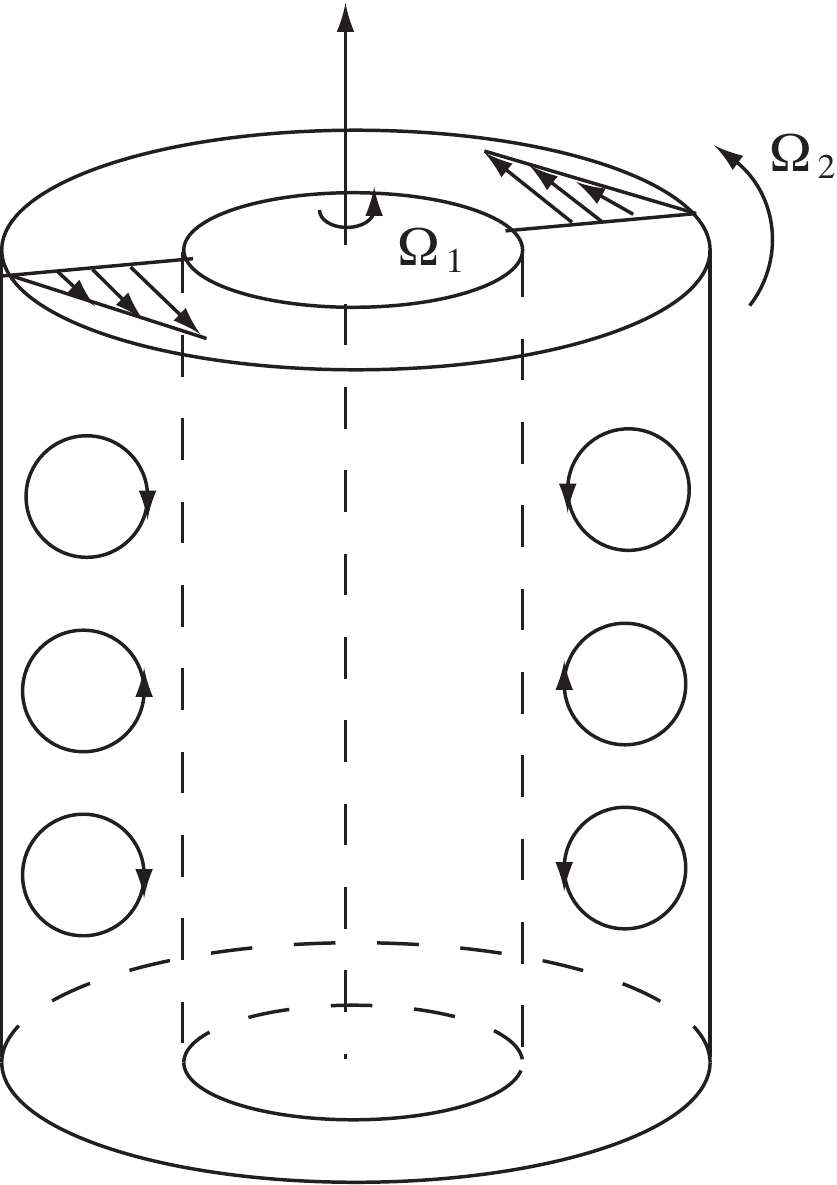}
  \caption{Couette Flow and Taylor vortices}\la{f9.8}
 \end{SCfigure}

There have been extensive studies for the Taylor problem from both the mathematical and physical point of view; see among many others, \cite{yudovich66, velte, chandrasekhar, dr}. Over the years, the Taylor problem, together with the Rayleigh-B\'enard convection problem, has
become one of the paradigms for studying nonequilibrium phase transitions and pattern formation in nonlinear sciences. 
 
The main objective of this article is to address the dynamic transition of the Taylor-Couette flow, and and study the formation and stability in its structure of the Taylor vortices. The main technical tools are the dynamical transition theory and the geometric theory for incompressible flows, both developed recently by the authors; see \cite{amsbook, ptd} and the references therein. 

The main philosophy of the dynamic transition theory is to search for  the full set of  transition states, giving a complete characterization on stability and  transition. The set of transition states is represented by a local attractor. Following this philosophy, the dynamic transition theory is developed  to identify the transition states and to classify them both dynamically and physically. One important ingredient of this theory is the introduction of a dynamic classification scheme of phase transitions. With this classification scheme, phase transitions are classified into three types: continuous (Type-I), jump (Type-II) and mixed (Type-III). 
The dynamic transition theory is recently developed by the authors   to identify the transition states and to classify them both dynamically and physically; see above references for details. 
The theory is motivated by phase transition problems in nonlinear sciences. Namely, the mathematical theory is developed under close links to the physics, and in return the theory is applied to the physical problems, although more applications are yet to be explored. With this theory, many long standing phase transition problems are either solved or become more accessible, providing new insights to  both theoretical and experimental studies for the underlying physical problems.

For simplicity, we focus in this article on  the $z$-periodic boundary
condition, which  is an approximate description for the case where the
ratio $L/(r_2-r_1)$ between the height $L$ and the gap $r_2-r_1$ is
sufficiently large.  We remark that similar results hold true as well for other type of boundary conditions, as well as for three dimensional perturbations (in the narrow-gap case); we refer the interested readers to \cite{ptd} for further details.

\medskip

The main results obtained are as follows.

{\sc First}, we show that the system always undergoes a dynamic transition as the Taylor number $T$ crosses the critical Taylor number $T_c$. The types of the transition can be either continuous (Type-I) or jump (Type-II), and are dictated precisely by the sign
of a nondimensional parameter $R$,  given completely by the first
eigenvectors, the ratio of the angular velocity of the outer and inner cylinders $\mu$,  and and the ratio of the radii of the inner and outer cylinders $\eta$.

\medskip

{\sc Second}, when $R<0$,  the transition is continuous, and  the critical
exponent of the phase transition, i.e., the exponent in the expression of
bifurcated solutions, is $\beta =1/2$. Moreover, there is
only one critical Taylor number $T_c$ such that the secondary  flow tends to the basic flow (Couette flow)
 as $T\rightarrow T_c$.

Also, for the narrow-gap case, the parameter $R$ defined by
(\ref{9.167}) is negative: $R<0$,  provided the two coaxial cylinders
rotating in the same direction, including the case where  the outer cylinder does not rotate.

\medskip

{\sc Third}, when $R>0$, the transition is a jump transition,  leading to more drastic changes,  coexistence of metastable states, and potentially more chaotic/turbulent behavior. In particular, 
there are two critical Taylor numbers $T_c$ and $T^*$ with $T^*<T_c$. 
When $T^*<T<T_c$,   the system has  two metastable states $\Sigma_0$, the trivial Couette flow 
and $\Sigma^T$, a local attractor away from the Couette flow. When $T > T_c$, the solution always moves away from the basic Couette flow to a more chaotic/turbulent regime, represented by the local attractor $\Sigma^T$. 

\medskip

{\sc Fourth},  the theoretic analysis carried out in this article shows that a street  of vortices
appear in the secondary flow for the narrow-gap case with
$\mu\rightarrow 1$. Thus the theoretic results are in agreement with
the Taylor experiments.

\bigskip

The article is organized as follows. The partial differential equation model and the set-up are given in Section 2, and the main dynamic transition theorems are given in Section 3. Explicit expressions of the parameter $R$ for determining the types of transitions  are further discussed in Section 4. The formation and structural stability of the Taylor vortices are addressed further in Section 5, and the main theorems are proved in Section 6.

\section{The Taylor Problem}
\subsection{Couette flow and Taylor vortices}
Consider an incompressible viscous fluid between two coaxial
cylinders. Let $r_1$ and $r_2$  $(r_2>r_1)$ be the radii of the two
cylinders, $\Omega_1$ and $\Omega_2$ the angular velocities of the
inner and the outer cylinders respectively, and
\begin{equation}
\mu =\Omega_2/\Omega_1,\ \ \ \ \eta =r_1/r_2.\label{9.96}
\end{equation}

The nondimensional Taylor number is defined by 
\begin{equation}
T=\frac{4h^4\Omega^2_1}{\nu^2},\label{9.97}
\end{equation}
where $\nu>0$ is the kinematic viscosity,   and  $h$  is  the vertical length scale.

There exists a basic steady state flow, called the Couette flow. In
the cylindrical polar coordinate $(r,\theta ,z)$, the Couette flow
is defined by
\begin{equation}
\left.
\begin{aligned} &(u_r,u_{\theta},u_z, p)=\left(0,V(r),0,\rho\int\frac{1}{r}V^2(r)dr\right),\\
&V(r)=ar+b/r,
\end{aligned}
\right.\label{9.98}
\end{equation}
where $(u_r,u_{\theta},u_z)$ is the velocity field, $p$  is  the pressure,
and $a,b$ are constants. It follows from the boundary conditions
that
$$V(r_1)=\Omega_1r_1,\ \ \ \ V(r_2)=\Omega_2r_2,$$
and the constants a and $b$ in (\ref{9.98}) are given by
$$a=-\Omega_1\eta^2\frac{1-\mu /\eta^2}{1-\eta^2},\ \ \ \
b=\Omega_1\frac{r^2_1(1-\mu )}{1-\eta^2}.$$ where $\mu$ and $\eta$
are given by (\ref{9.96}).

Based on the Rayleigh criterion, when $\mu >\eta^2$, the Couette
flow is always stable at a distribution of angular velocities
$$\Omega (r)=a+b/r^2\qtf  r_1<r<r_2.$$
However, when $\mu <\eta^2$,  the situation is different. As in the
Taylor experiments, consider  the case where the gap $r_2-r_1$ is much
smaller than the mean radius $r_0=1/2(r_1+r_2)$, namely,
$$r_2-r_1\ll  (r_1+r_2)/2,$$
and the two cylinders rotate in the same direction.  If the Taylor
number $T$ in (\ref{9.97}) satisfies  $T<T_c$, then the Couette flow (\ref{9.98})
is stable, and if  $T_c<T<T_c+\varepsilon$ for some $\varepsilon
>0$, a street  of vortices along the $z$-axis, called the Taylor
vortices, emerge  abruptly from the basic flow, as shown in Figure
\ref{f9.8}, and the corresponding flow pattern  is radically symmetric and structurally stable.

When the gap $r_2-r_2$ is not small than $r_0=1/2(r_1+r_2)$,
or  when the cylinders rotate in the opposite directions, the phenomena
one observes are much more complex; see  \cite{chandrasekhar} for
details.

Hence, in this section we always assume the condition
\begin{equation}
\eta^2>\mu\geq 0.\label{9.99}
\end{equation}

\subsection{Governing equations}
The hydrodynamic equations governing an incompressible viscous fluid
between two coaxial cylinders are the Navier-Stokes equations. In
the cylindrical polar coordinates $(r,\theta ,z)$, they are given by
\begin{equation}
\begin{aligned} 
&\frac{\partial u_r}{\partial t}+(u\cdot\nabla
)u_r-\frac{u^2_{\theta}}{r}=\nu \left(\Delta
u_r-\frac{2}{r^2}\frac{\partial
u_{\theta}}{\partial\theta}-\frac{u_r}{r^2}\right)-\frac{1}{\rho}\frac{\partial
p}{\partial r},\\
&\frac{\partial u_{\theta}}{\partial t}+(u\cdot\nabla
)u_{\theta}+\frac{u_ru_{\theta}}{r}=\nu \left(\Delta
u_{\theta}+\frac{2}{r^2}\frac{\partial
u_r}{\partial\theta}-\frac{u_{\theta}}{r^2}\right)-\frac{1}{r\rho}\frac{\partial
p}{\partial\theta},\\
&\frac{\partial u_z}{\partial t}+(u\cdot\nabla )u_z=\nu\Delta
u_z-\frac{1}{\rho}\frac{\partial p}{\partial z},\\
&\frac{\partial (ru_r)}{\partial r}+\frac{\partial
u_{\theta}}{\partial\theta}+\frac{\partial (ru_z)}{\partial z}=0,
\end{aligned}
\label{9.100}
\end{equation}
where $\nu$ is the kinematic viscosity, $\rho$  is  the density,
$u=(u_r,u_{\theta},u_z)$ is the velocity field, $p$  is  the pressure
function, and
\begin{align*}
& u\cdot\nabla= u_r\frac{\partial}{\partial
r}+\frac{u_{\theta}}{r}\frac{\partial}{\partial\theta}+u_z\frac{\partial}{\partial
z},\\
& \Delta=\frac{\partial^2}{\partial
r^2}+\frac{1}{r}\frac{\partial}{\partial
r}+\frac{1}{r^2}\frac{\partial^2}{\partial\theta^2}+\frac{\partial^2}{\partial
z^2}.
\end{align*}

Then it is easy to see that the Couette flow (\ref{9.98}) is a
steady state solution of (\ref{9.100}). In order to investigate its
stability and  transitions, we need to consider the perturbed
state of (\ref{9.98}):
$$u_r,u_{\theta}+V(r),u_z, p+\rho\int\frac{1}{r}v^2(r)dr.$$
The perturbed equations read
\begin{equation}
\left.
\begin{aligned}
& 
\frac{\partial u_r}{\partial t}+(u\cdot\nabla)u_r-\frac{u^2_{\theta}}{r}
 = \nu \left(\Delta u_r-\frac{2}{r^2}\frac{\partial
u_{\theta}}{\partial\theta}-\frac{u_r}{r^2}\right)-\frac{1}{\rho}\frac{\partial
p}{\partial r}\\
& \qquad \qquad \qquad 
   +\frac{2V(r)}{r}u_{\theta}  -  \frac{V(r)}{r}\frac{\partial
u_r}{\partial\theta},
\\
&\frac{\partial u_{\theta}}{\partial t}+(u\cdot\nabla)u_{\theta}+\frac{u_{\theta}u_r}{r}
 =\nu\left(\Delta u_{\theta}+\frac{2}{r^2}\frac{\partial
u_r}{\partial\theta}-\frac{u_{\theta}}{r^2}\right)\\
&\qquad \qquad \qquad  
  -\frac{1}{r\rho}\frac{\partial p}{\partial\theta}-(V^{\prime}
    +\frac{V}{r})u_r-\frac{V}{r}\frac{\partial
u_{\theta}}{\partial\theta}, \\
&   \frac{\partial
u_z}{\partial t}+(u\cdot\nabla )u_z=\nu\Delta
u_z-\frac{1}{\rho}\frac{\partial p}{\partial
z}-\frac{V}{r}\frac{\partial u_z}{\partial\theta},  \\
&\frac{\partial (ru_z)}{\partial z}+\frac{\partial (ru_r)}{\partial
r}+\frac{\partial u_{\theta}}{\partial\theta}=0. 
\end{aligned}\right.
\label{9.101}
\end{equation}

To derive 
the nondimensional form of equations (\ref{9.101}), let
\begin{align*}
&(x, t) =( hx^{\prime},  h^2t^{\prime}/\nu) && (x=(r,r\theta ,z)),\\
&(u, p)=( \nu u^{\prime}/h, \rho\nu^2p^{\prime}/h^2) &&  (u=(u_r,u_{\theta},u_z)).
\end{align*}
Omitting the primes, we obtain the nondimensional form of
(\ref{9.101}) as follows:
\begin{equation}\label{9.102}
\left.
\begin{aligned}
&
\frac{\partial u_r}{\partial t} 
= \Delta u_r-\frac{2}{r^2}\frac{\partial u_{\theta}}{\partial\theta}-\frac{u_r}{r^2}
  -(u\cdot\nabla )u_r+\frac{u^2_{\theta}}{r}-\frac{\partial p}{\partial r}\\
& 
   \qquad \qquad 
-\sqrt{T}\left(\frac{\eta^2-\mu}{1-\eta^2}-\frac{1-\mu}{1-\eta^2}\frac{r^2_1}{r^2}\right)\left(
u_{\theta}-1/2\frac{\partial u_r}{\partial\theta}\right),\\
&
\frac{\partial u_{\theta}}{\partial t} = \Delta
u_{\theta}+\frac{2}{r^2}\frac{\partial
u_r}{\partial\theta}-\frac{u_{\theta}}{r^2}-(u\cdot\nabla
)u_{\theta}-\frac{u_{\theta}u_r}{r}-\frac{1}{r}\frac{\partial
p}{\partial\theta}\\
&   
 \qquad \qquad
  +\sqrt{T}\frac{\eta^2-\mu}{1-\eta^2}u_r+\frac{\sqrt{T}}{2}\left(\frac{\eta^2-\mu}{1-\eta^2}
-\frac{1-\mu}{1-\eta^2}\frac{r^2_1}{r^2}\right)\frac{\partial
u_{\theta}}{\partial\theta},\\
&
\frac{\partial u_z}{\partial t}=\Delta u_z-(u\cdot\nabla
)u_z-\frac{\partial p}{\partial
z}+\frac{\sqrt{T}}{2}\left(\frac{\eta^2-\mu}{1-\eta^2}-\frac{1-\mu}{1-\eta^2}
\frac{r^2_1}{r^2}\right)\frac{\partial u_z}{\partial\theta}, \\
&\frac{\partial (ru_z)}{\partial z}+\frac{\partial (ru_r)}{\partial
r}+\frac{\partial u_{\theta}}{\partial\theta}=0,
\end{aligned}\right.
\end{equation}
where $T$ is the Taylor number as defined in (\ref{9.97}).

The nondimensional domain for (\ref{9.102}) is
$$\Omega =(l_1,l_2)\times (0,2\pi )\times (0,L),$$
where $l_i=r_i/h$  $(i=1,2)$,  and $L$ is the height of the fluid between the
two cylinders. The initial value condition for (\ref{9.102}) is
given by \begin{equation} u(r,\theta ,z,0)=u_0(r,\theta
,z).\label{9.103}
\end{equation}
There are different physically sound boundary conditions. In the
$\theta$-direction it is periodic
\begin{equation}
u(r,\theta +2k\pi ,z)=u(r,\theta ,z),\ \ \ \ \forall
k\in\mathbb{Z}.\label{9.104}
\end{equation}
In the radical direction, there is the rigid boundary condition
\begin{equation}
u=(u_z,u_r,u_{\theta})=0,\ \ \ \ \text{at}\ \ \ \
r=l_1,l_2.\label{9.105}
\end{equation}
At the top and bottom in the $z$-direction $(z=0,L)$, either the
free boundary condition or the rigid boundary condition or the
periodic boundary condition can be used:

{\sc Dirichlet Boundary Condition:}
\begin{equation}
u=(u_r,u_{\theta},u_z)=0\ \ \ \ \text{at}\ \ \ \ z=0,L;\label{9.106}
\end{equation}

{\sc Free-Slip Boundary Condition:}
\begin{equation}
u_z=0,\ \ \ \ \frac{\partial u_r}{\partial z}=\frac{\partial
u_{\theta}}{\partial z}=0\ \ \ \ \text{at}\ \ \ \ z=0,L;\label{9.107}
\end{equation}

{\sc Free-Rigid Boundary Condition:}
\begin{equation}
\left.
\begin{aligned} 
& u_z=0,\frac{\partial u_r}{\partial
z}=\frac{\partial u_{\theta}}{\partial z}=0&&\text{at}\ z=L,\\
& u=(u_z,u_r,u_{\theta})=0 &&\text{at}\ z=0;
\end{aligned}
\right.\label{9.108}
\end{equation}

{\sc Periodic Boundary Condition:}
\begin{equation}
u(r,\theta ,z+2kL)=u(r,\theta z) \ \ \ \ \forall
k\in\mathbb{Z}.\label{9.109}
\end{equation}

\section{Dynamic Transitions}

\subsection{Functional setting}
We now study the Taylor problem (\ref{9.102}) with the $z$-periodic
boundary condition (\ref{9.109})  and with  axisymmetric perturbations.
Assuming that the equations (\ref{9.102}) are independent of
$\theta$, and taking the length scale $h=r_2$ in the nondimensional
form,  we obtain
\begin{eqnarray}
\left.\begin{aligned} &\frac{\partial u_z}{\partial t}=\Delta
u_z-\frac{\partial p}{\partial z}-(\tilde{u}\cdot\nabla )u_z,\\
&\frac{\partial u_r}{\partial t}=(\Delta -\frac{1}{r^2})u_r+\lambda
\left(\frac{1}{r^2}-\kappa\right)u_{\theta}-\frac{\partial
p}{\partial r}+\frac{u^2_{\theta}}{r}-(\tilde{u}\cdot\nabla )u_r,\\
&\frac{\partial u_{\theta}}{\partial t}=\left(\Delta
-\frac{1}{r^2}\right)u_{\theta}+\lambda\kappa
u_r-\frac{u_ru_{\theta}}{r}-(\tilde{u}\cdot\nabla )u_{\theta},\\
&\frac{\partial (ru_z)}{\partial z}+\frac{\partial (ru_r)}{\partial
r}=0,
\end{aligned}
\right.\label{9.145}
\end{eqnarray}
where $\lambda =\sqrt{T}$,  $T$ is the Taylor number, and  
\begin{align*}
& T=\frac{4r^4_2\Omega^2_1(1-\mu )^2\eta^4}{\nu^2(1-\eta^2)^2},
&& \eta^2=r^2_1/r^2_2, \\
& \kappa =\frac{1-\mu /\eta^2}{1-\mu },&&
\Delta =\frac{\partial^2}{\partial
r^2}+\frac{1}{r}\frac{\partial}{\partial
r}+\frac{\partial^2}{\partial z^2}, \\
& (\widetilde u \cdot \nabla ) = u_r\, \frac{\partial }{\partial r} +
u_z\, \frac{\partial }{\partial z}.
\end{align*}

The nondimensional domain is $M=(\eta ,1)\times (0,L)$, 
and the boundary conditions take (\ref{9.105}) and
(\ref{9.109}), i.e., 
\begin{equation}
\left.
\begin{aligned} 
&u=(u_z,u_r,u_{\theta})=0  \qquad \text{at}\ r=\eta,1,\\
&u\ \text{is\ periodic\ with\ period}\ L\ \text{in\ the}\
z-\text{direction}.
\end{aligned}
\right.\label{9.146}
\end{equation}
The initial value condition is
\begin{equation}
u=u_0(r,z)\ \ \ \ \text{at}\ \ \ \ t=0.\label{9.147}
\end{equation}

For the Taylor problem (\ref{9.145})-(\ref{9.147}), we set
\begin{eqnarray*}
&&H=\left\{u=(\tilde{u},u_{\theta})\in
L^2(M)^3 \left|\begin{array}{l} \text{div}(r\tilde{u})=0,u_r=0\
\text{at}\ r=\eta ,1,\ \text{and}\\
u\ \text{is}\ L-\text{periodic\ in\ the}\ z-\text{direction}
\end{array}
\right.\right\}\\
&&H_1=\left\{u\in H^2(M)^3 \cap H\ |\ u\ \text{satisfies}\
(\ref{9.146})\right\},
\end{eqnarray*}
and the inner product of $H$ is defined by
$$(u,v)_H=\int_Mu\cdot vrdzdr.$$

Let the linear operator $L_{\lambda}=-A+\lambda B:H_1\rightarrow H$
and nonlinear operator $G:H_1\rightarrow H$ be defined by
\begin{equation}
\left.\begin{aligned} &Au=-P\left(\Delta u_z,\left(\Delta
-\frac{1}{r^2}\right)u_r,\left(\Delta
-\frac{1}{r^2}\right)u_{\theta}\right),\\
&Bu=P\left(0,\left(\frac{1}{r^2}-\kappa\right)u_{\theta},\kappa
u_r\right),\\
&G(u)=-P\left((\tilde{u}\cdot\nabla )u_z,(\tilde{u}\cdot\nabla
)u_r-\frac{u^2_{\theta}}{r},(\tilde{u}\cdot\nabla
)u_{\theta}+\frac{u_{\theta}u_r}{r}\right), \end{aligned}
\right.\label{9.148}
\end{equation}
where $P:L^2(M)^3 \rightarrow H$ is the Leray projection. Thus the
Taylor problem (\ref{9.145})-(\ref{9.147}) is rewritten in the
abstract form
\begin{equation}
\left.
\begin{aligned} 
&\frac{du}{dt}=L_{\lambda}u+G(u),\\
&u(0)=u_0.
\end{aligned}
\right.\label{9.149}
\end{equation}

For simplicity, let  $G:H_1\rightarrow H$ be the corresponding  bilinear operator defined
by
$$G(u,v)=-P\left((\tilde{u}\cdot\nabla )v_z,(\tilde{u}\cdot\nabla
)u_r-\frac{u_{\theta}v_{\theta}}{r},(\tilde{u}\cdot\nabla
)v_{\theta}+\frac{u_{\theta}v_r}{r}\right).$$
Then it is easy to see that 
\begin{equation}
(G(u,v),w)_H=- (G(u,w),v)_H.\label{9.150}
\end{equation}

\subsection{Eigenvalue problem}

\medskip

To study the phase transition of the Taylor problem
(\ref{9.145})-(\ref{9.147}) it is necessary to consider the
eigenvalue problem of its linearized equation.
The associated
eigenvalue equation of (\ref{9.149}) is as follows:
\begin{equation}
L_{\lambda}u=-Au+\lambda Bu=\beta (\lambda )u,\label{9.151}
\end{equation}
and the conjugate equation of (\ref{9.151}) is given by
\begin{equation}
L^*_{\lambda}u^*=-A^*u^*+\lambda B^*u^*=\beta (\lambda
)u^*.\label{9.152}
\end{equation}

The equations corresponding to (\ref{9.151}) are as follows
\begin{equation}
\left.\begin{aligned} &\Delta u_z-\frac{\partial p}{\partial
z}=\beta (\lambda )u_z,\\
&\left(\Delta
-\frac{1}{r^2}\right)u_r+\lambda\left(\frac{1}{r^2}-\kappa\right)u_{\theta}-\frac{\partial
p}{\partial r}=\beta (\lambda )u_r,\\
&\left(\Delta -\frac{1}{r^2}\right)u_{\theta}+\lambda\kappa
u_r=\beta (\lambda )u_{\theta},\\
&\text{div}(r\tilde{u})=0.
\end{aligned}
\right.\label{9.153}
\end{equation}
The equations corresponding to (\ref{9.152}) are given by 
\begin{equation}
\left.\begin{aligned} 
&\Delta u^*_z-\frac{\partial p^*}{\partial
z}=\beta (\lambda )u^*_z,\\
&\left(\Delta -\frac{1}{r^2}\right)u^*_r+\lambda\kappa
u^*_{\theta}-\frac{\partial p^*}{\partial r}=\beta (\lambda
)u^*_r,\\
&\left(\Delta
-\frac{1}{r^2}\right)u^*_{\theta}+\lambda\left(\frac{1}{r^2}-\kappa\right)u^*_r=\beta
(\lambda )u^*_{\theta},\\
&\text{div}(r\tilde{u}^*)=0.
\end{aligned}
\right.\label{9.154}
\end{equation}
Both (\ref{9.153}) and (\ref{9.154}) are supplemented with the
boundary condition (\ref{9.146}).

We start with the principle of exchange of stability (PES). 
It is known that for each given period $L$, there is a
$\lambda^*_0=\lambda_0(L)$ such that the eigenvalues
$\beta_j(\lambda )  $   $(j=1,2,\cdots )$ of (\ref{9.153}) with
(\ref{9.146}) near $\lambda =\lambda^*_0$ satisfy that
$\beta_1(\lambda ),\cdots,\beta_m(\lambda )$  $(m\geq 1)$ are real, and
\begin{equation}
\left.
\begin{aligned} 
&
\beta_i(\lambda )
\left\{\begin{aligned}
 &  <0   &&\text{if }\ \lambda <\lambda^*_0,\\
 &  =0    &&\text{if }\ \lambda =\lambda^*_0
\end{aligned}
\right.   &&  \text{for } 1\leq i\leq m,\\
&
\text{Re}\beta_j(\lambda^*_0)<0  &&  \text{for }  j\geq m+1.
\end{aligned}
\right.\label{9.155}
\end{equation}
In addition, there is a period $L^{\prime}>0$ such that
\begin{equation}
\lambda_0=\lambda_0(L^{\prime})=\min\limits_{L>0}\lambda_0(L).\label{9.156}
\end{equation}
Thanks to \cite{yudovich66,velte}, for $\mu
=\Omega_1/\Omega_2\geq 0$ the multiplicity $m=2$ in (\ref{9.155}) at
$\lambda_0=\lambda_0(L^{\prime})$; see also \cite{kirch,temam84}.

In this section, we always take $L^{\prime}$ as the period given by (\ref{9.156}), and
define the following number as the critical Taylor number:
$$T_c=\lambda^2_0(L^{\prime}).$$
For simplicity, omitting the prime we denote $L^{\prime}$ by $L$.

By (\ref{9.155}) and (\ref{9.156}), to verify the PES it suffices to
prove that  for $\lambda >\lambda_0$, 
\begin{equation}
\beta_i(\lambda )>0\qquad \forall  1\leq i\leq m.\label{9.157}
\end{equation}
To this end,  we need to derive the
eigenvectors of (\ref{9.153}) and (\ref{9.154}) at
$\beta_i(\lambda_0)=0$   $(i=1,2).$

It is readily to check that the eigenvectors of (\ref{9.153}) with
(\ref{9.146}) corresponding to $\beta_i(\lambda_0)=0$  $(i=1,2)$ are
given by
\begin{align}
& \psi_1
= (\psi_z,  \psi_r,  \psi_\theta)  =( -\sin az \ D_*h(r),
a\cos az \ h(r),
\cos az \  \varphi (r)), \label{9.158}
\\
& \tilde{\psi}_1
  =(\tilde\psi_z, \tilde \psi_r, \tilde \psi_\theta) 
   =(\cos az \  D_*h(r),
a\sin az \  h(r),
\sin az \  \varphi (r)),
\label{9.159}
\end{align}
where $(h(r),\varphi (r))$ satisfies
\begin{equation}
\left.
\begin{aligned}
&(DD_*-a^2)^2h=a^2\lambda_0\left(\frac{1}{r^2}-\kappa\right)\varphi,\\
&(DD_*-a^2)\varphi =-\lambda_0\kappa h,\\
&  (h, Dh,  \varphi)  =0  &&  \text{at }  r=\eta ,1,
\end{aligned}
\right.\label{9.160}
\end{equation}
and
$$D=\frac{d}{dr},\ \ \ \ D_*=\frac{d}{dr}+\frac{1}{r},\ \ \ \
a=\frac{2\pi}{L}.$$ 
The dual eigenvectors of (\ref{9.154}) with
(\ref{9.146}) read
\begin{align}
&
\psi^*_1=(\psi^*_z,  \psi^*_r, \psi^*_{\theta}) 
=( -\sin az \ D_*h^*(r), a\cos az\  h^*(r), \cos az \  \varphi^*(r)), \label{9.161}
\\
&
\tilde{\psi}^*_1=(\tilde{\psi}^*_z, \tilde{\psi}^*_r, \tilde{\psi}^*_{\theta})
=( \cos az  \ D_*h^*(r), a\sin az \ h^*(r), \sin az   \ \varphi^*(r)), \label{9.162}
\end{align}
where $(h^*,\varphi^*)$ satisfies that
\begin{equation}
\left.
\begin{aligned} 
&(DD_*-a^2)^2h^*=\lambda_0\kappa\varphi^*,\\
&(DD_*-a^2)\varphi^*=-a^2\lambda_0\left(\frac{1}{r^2}-\kappa\right)h^*,\\
&(h^*, Dh^*,  \varphi^*)=0 &&  \text{at  }  r=\eta ,1.
\end{aligned}
\right.\label{9.163}
\end{equation}

The following lemma shows that the PES is valid for the Taylor
problem (\ref{9.145})-(\ref{9.147}) with  $\mu\geq 0$.

\bl\la{l9.4}
 If $\mu\geq 0$, then the first eigenvalues
$\beta_i(\lambda )$  $(1\leq i\leq m)$ of (\ref{9.153}) are real with
multiplicity $m=2$ near $\lambda =\lambda_0 = \sqrt{T_c}$, and
the first eigenvectors at $\lambda =\lambda_0$ are given by
(\ref{9.158}) and (\ref{9.159}). Moreover, the eigenvalues
$\beta_j(\lambda )$  $(j=1,2,\cdots )$ satisfy the conditions (5.4) and
(5.5) at $\lambda =\lambda_0$, i.e.,  the PES
holds true at the critical Taylor number $T_c$.
\el

\bp
We only need to prove (\ref{9.157}). By Theorem~2.1 in \cite{MW08k}, it suffices to verify that
\begin{equation}
(B\psi_1,\psi^*_1)_H\neq 0,  \ \ \ \
(B\tilde{\psi}_1,\tilde{\psi}^*_1)_H\neq 0.\label{9.164}
\end{equation}

We infer from (\ref{9.148}), (\ref{9.158}), (\ref{9.159}),
(\ref{9.161}), and (\ref{9.162}) that
\begin{eqnarray}
(B\psi_1,\psi^*_1)_H&=&(B\tilde{\psi}_1,\tilde{\psi}^*_1)_H\label{9.165}\\
&=&\int^L_0\int^1_{\eta}r\left[\left(\frac{1}{r^2}-\kappa\right)\psi_{\theta}\psi^*_r+\kappa\psi_r
\psi^*_{\theta}\right]dzdr\nonumber\\
&=&\frac{La}{2}\int^1_{\eta}r\left[\left(\frac{1}{r^2}-\kappa\right)h\varphi^*+\kappa\varphi
h^*\right]dr.\nonumber
\end{eqnarray}
Since $\mu\geq 0$, by (\ref{9.99}),  we have $0<\kappa <1$ and
$\frac{1}{r^2}-\kappa >0$ for $\eta <r<1$. On the other hand, we
know that the first eigenvectors $(h(r),\varphi (r))$ of
(\ref{9.160}) and $(h^*(r),\varphi^*(r))$ of (\ref{9.163}) at
$\lambda =\lambda_0$ are positive; see  \cite{velte,kirch,temam84}:
\begin{equation}
h(r)>0,\ \ \ \ \varphi (r)>0,\ \ \ \ h^*(r)>0,\ \ \ \
\varphi^*(r)>0 \ \ \ \ \forall\eta <r<1.\label{9.166}
\end{equation}
Thus (\ref{9.164}) follows from (\ref{9.165}) and (\ref{9.166}).
The proof is complete.
\ep

\subsection{Phase transition theorems}
Here  we always assume that the first eigenvalue of (\ref{9.153})
with (\ref{9.146}) is real with multiplicity $m=2$, i.e.,  the first
eigenvalue $\lambda_0$ of (\ref{9.160}) is simple, and the PES holds
true. By Lemma \ref{l9.4}, this assumption is valid for all $\mu\geq 0$ and
$0<\eta <1.$

Let $\psi_1$ and $\psi^*_1$ be given by (\ref{9.158}) and
(\ref{9.161}). We define a number $R$ by
\begin{equation}
R=\frac{1}{(\psi_1,\psi^*_1)_H}\left[(G(\Phi
,\psi_1),\psi^*_1)_H+(G(\psi_1,\Phi
),\psi^*_1)_H\right],\label{9.167}
\end{equation}
where $\Phi \in H_1$ is defined by
\begin{equation}
(A-\lambda_0B)\Phi =G(\psi_1,\psi_1).\label{9.168}
\end{equation}
Here the operator $A,B$ and $G$ are as in (\ref{9.148}). The
solution $\Phi$ of (\ref{9.168}) exists because $G(\psi ,\psi_1)$ is
orthogonal with $\psi^*_1$ and $\tilde{\psi}^*_1$ in $H$.

The following results characterize the dynamical properties of phase
transitions for the Taylor problem with the $z$-periodic boundary
condition.

\bt\la{t9.10}
 If the number $R<0$ in (\ref{9.167}), then the
Taylor problem (\ref{9.145})-(\ref{9.147}) has a Tyep-I (continuous)
transition at the critical Taylor number $T=T_c$ or $\lambda
=\lambda_0$, and the following assertions holds true:

\begin{itemize}

\item[(1)] When the Taylor number $T\leq T_c$ or
$\lambda\leq\lambda_0$, the steady state $u=0$ is locally
asymptotically stable.

\item[(2)] The problem bifurcates from $(u,\lambda
)=(0,\lambda_0)$ (or from $(u,T)=(0,T_c))$ to an attractor
${\mathcal{A}}_{\lambda}$ homeomorphic to a circle $S^1$ on
$\lambda_0<\lambda$, which consists of steady states of this
problem.

\item[(3)] Any $u\in{\mathcal{A}}_{\lambda}$ can be expressed
as \begin{eqnarray*} &&u=|\beta_1(\lambda
)/R|^{1/2}v+o(|\beta_1|^{1/2}),\\
&&v=x\psi_1+y\tilde{\psi}_1,\\
&& x^2+y^2=1,
\end{eqnarray*}
where $\psi_1,\tilde{\psi}_1$ are given in (\ref{9.158}) and
(\ref{9.161}).

\item[(4)] There is an open set $U\subset H$ with $0\in U$ such
that ${\mathcal{A}}_{\lambda}$ attracts $U \setminus \Gamma$, where $\Gamma$
is the stable manifold of $u=0$ with codimension two in $H$.

\item[(5)] When $1-\mu >0$ is small, for any $u_0\in
U \setminus (\Gamma\cup\widetilde{H})$, there exists a time $t_0\geq 0$ such that
for any $t>t_0$, the vector field $\tilde{u}(t,u_0)=(u_z,u_r)$ is
topologically equivalent to one of the patterns shown in Figure \ref{f9.12}, 
where $u=(\tilde{u}(t,u_0),u_{\theta}(t,u_0))$ is the solution of
(\ref{9.145})-(\ref{9.147}), and
$$\widetilde{H}=\{u=(u_z,u_r,u_{\theta})\in H|\
\int^L_0\int^1_{\eta}ru_zdrdz=0\}.$$

\item[(6)] When $1-\mu >0$ is small, for any $u_0\in
(U\cap\tilde{H}) \setminus \Gamma$, there exists a time $t_0\geq 0$ such that
for any $t>t_0, \tilde{u}(t,u_0)=(u_z,u_r)$ is topologically
equivalent to the structure as shown in Figure \ref{f9.13}.
\end{itemize}
\et

\begin{SCfigure}[25][t]
  \centering
  \includegraphics[width=.25\textwidth]{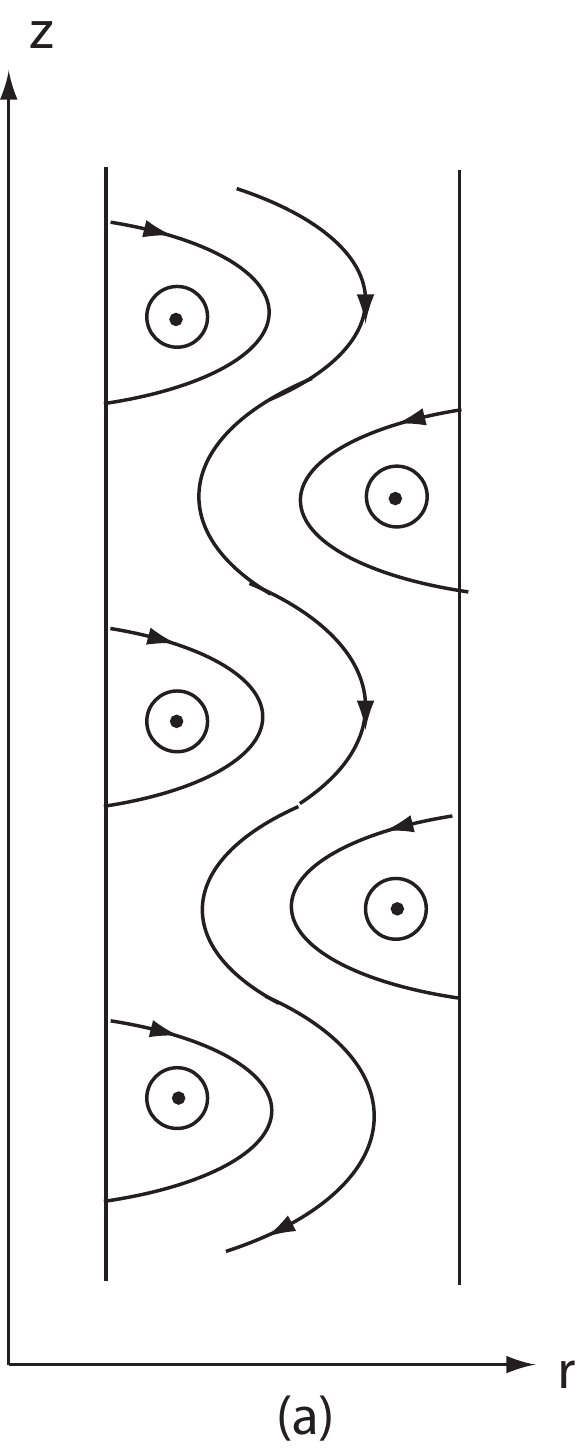} \qquad 
  \includegraphics[width=.25\textwidth]{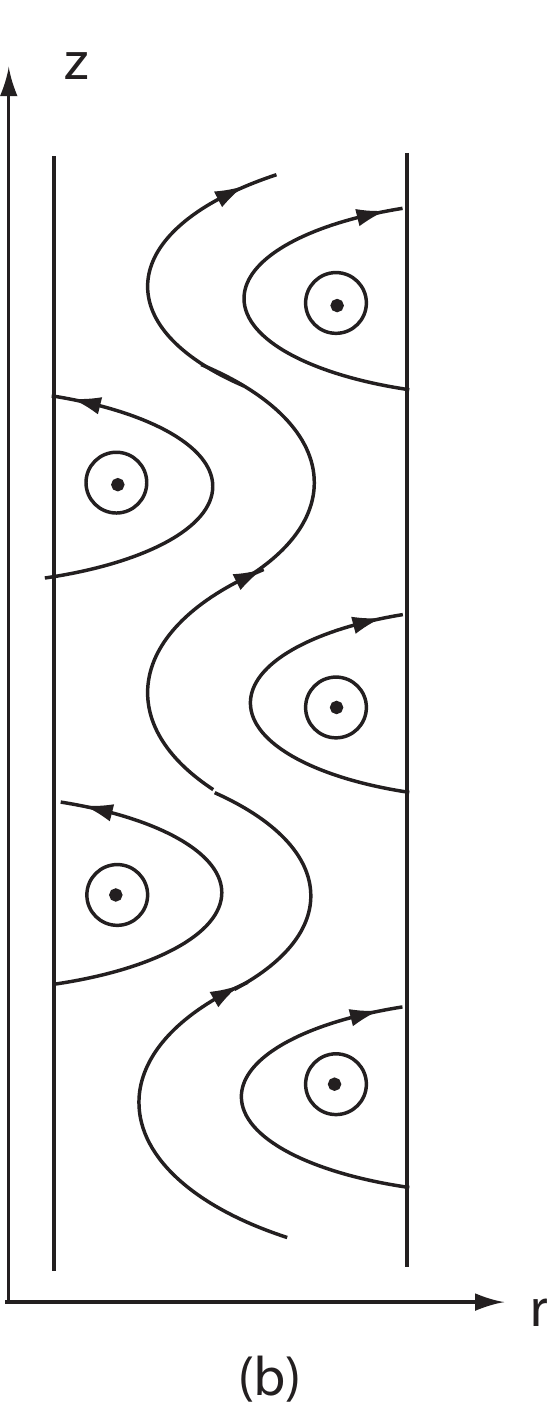}
  \caption{Taylor vortices with a cross-channel flow.}\la{f9.12}
 \end{SCfigure}

\begin{SCfigure}[25][t]
  \centering
  \includegraphics[width=.25\textwidth]{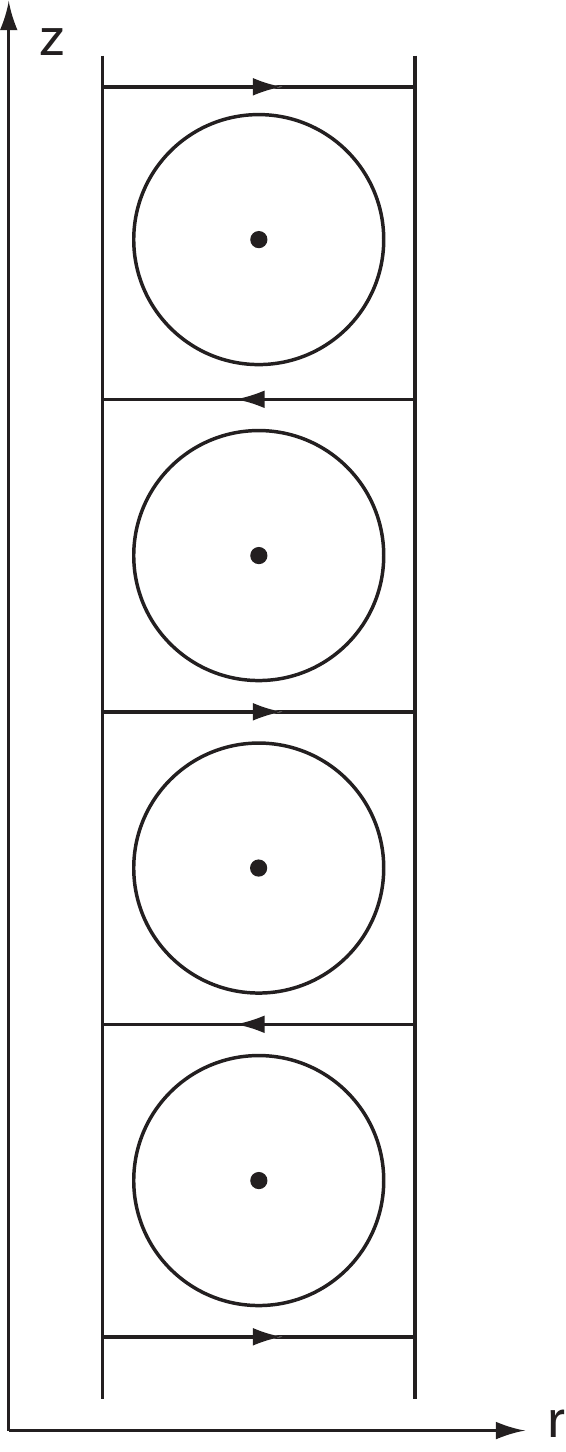}
    \caption{Taylor vortices  without a cross-channel flow.}\la{f9.13}
 \end{SCfigure}

\bt\la{t9.11}
 For the case where $R>0$, the transition of
the Taylor problem (\ref{9.145})-(\ref{9.147}) at $T=T_c$ is of Type-II. 
Moreover, the Taylor problem has a singularity separation
at $T^*<T_c$  $(\lambda^*<\lambda_0)$. More precisely we have the
following assertions:

\begin{itemize}
\item[(1)] There exists a number $\lambda^*$  $(0<\lambda^*<\lambda_0)$
such that the problem generate a circle $\Sigma^*=S^1$ at $\lambda
=\lambda^*$ consisting of singular points, and bifurcates from
$(\Sigma^*,\lambda^*)$ on $\lambda^*<\lambda$ to at least  two
branches of circles $\Sigma^{\lambda}_1$ and $\Sigma^{\lambda}_2$,
each consisting of steady states satisfying
\begin{align*}
&\lim_{\lambda \to \lambda_0}\Sigma^1_{\lambda}=\{0\},\\
&\text{dist}\left(\Sigma^{\lambda}_2,0\right)=\min_{u\in\Sigma^{\lambda}_2}\|u\|_H>0&&  \text{at}\ \ \ \ \lambda =\lambda_0;
\end{align*}
see Figure \ref{f9.14}.

\item[(2)] For each $\lambda^*<\lambda <\lambda_0$, the space $H$
can be decomposed into two open sets $U^{\lambda}_1$ and
$U^{\lambda}_2:H=\bar{U}^{\lambda}_1+\bar{U}^{\lambda}_2$ with
$U^{\lambda}_1\cap U^{\lambda}_2=\emptyset
,\Sigma^{\lambda}_1\subset\partial U^{\lambda}_1\cap\partial
U^{\lambda}_2$ such that the problem has two disjoint attractors
${\mathcal{A}}^{\lambda}_1$ and ${\mathcal{A}}^{\lambda}_2$:
$${\mathcal{A}}^{\lambda}_1=\{0\}\subset U^{\lambda}_1,\ \ \ \
\Sigma^{\lambda}_2\subset{\mathcal{A}}^{\lambda}_2\subset
U^{\lambda}_2,$$ and ${\mathcal{A}}^{\lambda}_i$ attracts
$U^{\lambda}_i$  $(i=1,2)$.

\item[(3)] For  $\lambda_0\leq\lambda$, the problem has an attractor
${\mathcal{A}}^{\lambda}$ satisfying
$$\lim\limits_{\lambda\rightarrow\lambda_0}{\mathcal{A}}^{\lambda}_2={\mathcal{A}}^{\lambda_0},\
\ \ \ \text{dist}({\mathcal{A}}^{\lambda},0)>0\ \ \ \
\forall\lambda\geq\lambda_0,$$ 
and ${\mathcal{A}}^{\lambda}$
attracts $H \setminus \Gamma_{\lambda}$, 
where $\Gamma_{\lambda}$ is the
stable manifold of $u=0$ with codimension $m_{\lambda}\geq 2$ in
$H$.
\end{itemize}
\et

\begin{SCfigure}[25][t]
  \centering
  \includegraphics[width=0.5\textwidth]{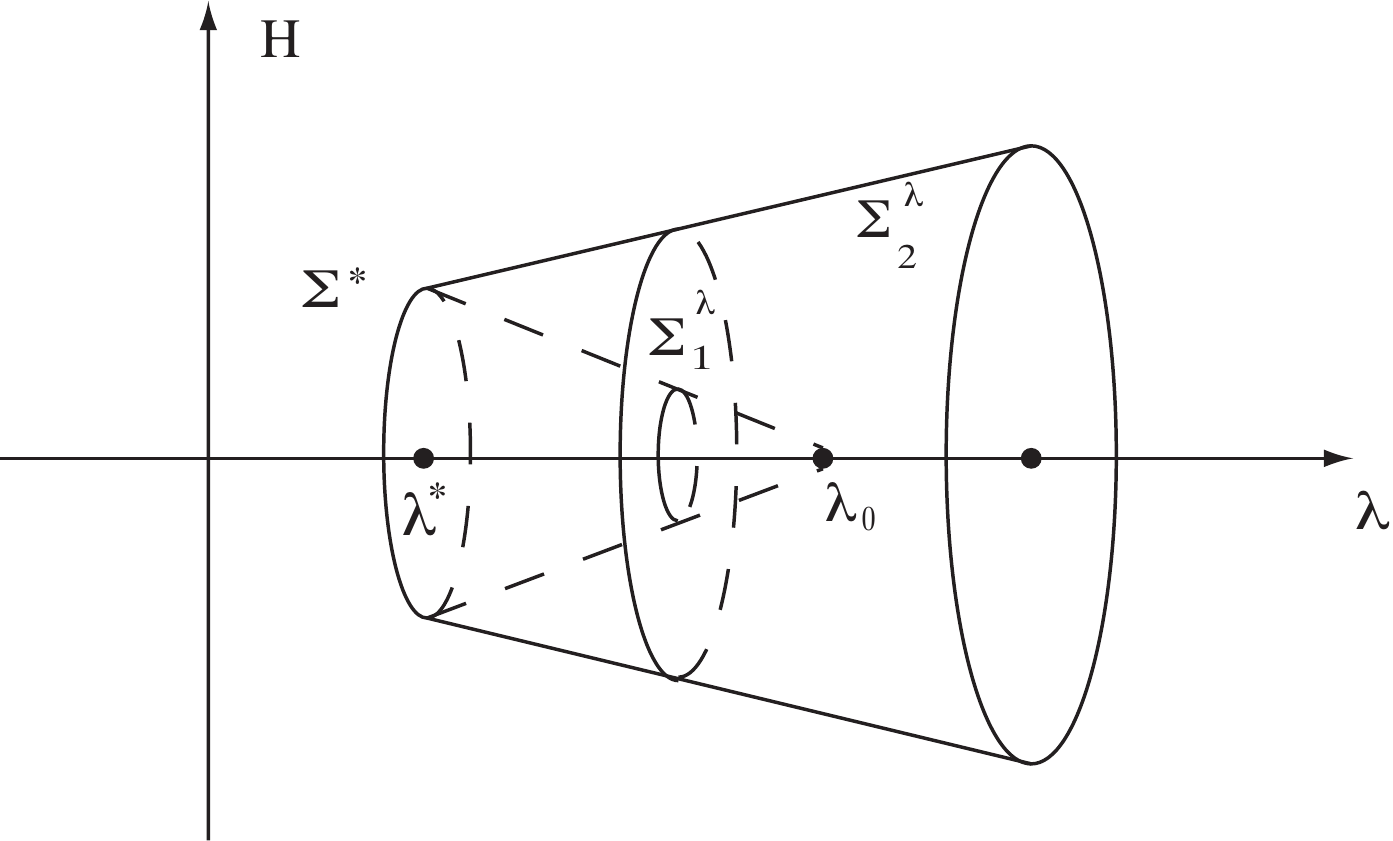}
  \caption{Singularity separation of circles consisting of
steady states at $\lambda =\lambda^*$.}\la{f9.14}
 \end{SCfigure}
 
\section{Explicit expression of the parameter $R$}
\subsection{General case}
The parameter $R$ defined by (\ref{9.167}) and (\ref{9.168}) can be
explicitly expressed in the following integral formula:
\begin{align*}
R=
&
-\frac{1}{(\psi_1,\psi^*_1)_H}
\Big[\frac{\pi}{2}\int^1_{\eta}rh\varphi^*\frac{d\phi_0}{dr}dr  \\
&
+\int^L_0\int^1_{\eta}r\Big((\tilde{\phi}\cdot\nabla)\psi_z\psi^*_z+(\tilde{\psi}\cdot\nabla
)\phi_z\psi^*_z+(\tilde{\phi}\cdot\nabla )\psi_r\psi^*_r  \\
&
+(\tilde{\psi}\cdot\nabla )\phi_r\psi^*_r+(\tilde{\phi}\cdot\nabla)\psi_{\theta}\psi^*_{\theta}
+(\tilde{\psi}\cdot\nabla)\phi_{\theta}\psi^*_{\theta}\\
&
+ \frac{\phi_{\theta}\psi_r\psi^*_{\theta}}{r}+\frac{\psi_{\theta}\phi_r\psi^*_{\theta}}{r}-
2\frac{\psi_{\theta}\phi_{\theta}\psi^*_r}{r}\Big)drdz\Big],
\end{align*}
where 
$\tilde{\psi}=(\psi_z,\psi_r),\psi_1=(\psi_z,\psi_r,\psi_{\theta}),\psi^*_1=(\psi^*_z,\psi^*_r,
\psi^*_{\theta})$ are given by (\ref{9.158}) and (\ref{9.161}), 
$$(\psi_1,\psi^*_1)_H=\int^L_0\int^1_{\eta}r(\psi_z\psi^*_z+\psi_r\psi^*_r+\psi_{\theta}\psi^*_{\theta})drdz,
$$
and
$\phi_0,\phi =(\phi_z,\phi_r,\phi_{\theta})$ satisfy
\begin{align*}
&
\left\{\begin{aligned} 
&DD_*\phi_0=a(\varphi D_*h+hD\varphi +\frac{1}{r}\varphi h),\\
&\phi_0|_{r=\eta, 1}=0 ,
\end{aligned}
\right.\\
&
\left\{\begin{aligned} 
&-\Delta\phi_z+\frac{\partial p}{\partial z}=-1/2\sin 2azH_1(r),\\
&-\left(\Delta
-\frac{1}{r^2}\right)\phi_r-\lambda_0\left(\frac{1}{r^2}-\kappa\right)\phi_{\theta}+\frac{\partial
p}{\partial r}=-1/2\cos 2azH_2(r),\\
&-\left(\Delta
-\frac{1}{r^2}\right)\phi_{\theta}-\lambda_0\kappa\phi_r=-1/2\cos
2azH_3(r),\\
&\text{div}(r\tilde{\phi})=0,\ \ \ \ \tilde{\phi}=(\phi_z,\phi_r),\\
&\phi |_{r=\eta, 1}=0.
\end{aligned}
\right.
\end{align*}
Here $H_i(r),i=1,2,3$, are as in (\ref{9.177}).

\subsection{Narrow-gap case}
We consider here the case where the gap $r_2-r_1$ is small compared
to the mean radius $r_0=(r_1+r_2)/2$ with $\mu\geq 0$   and with
axisymmetric perturbations. This case is the situation investigated
by  \cite{taylor} in 1923.

We take the length scale $h=r_2-r_1$. Then the narrow gap condition is given by 
\begin{equation}
1=r_2-r_1\ll (r_1+r_2)/2.\label{9.110}
\end{equation}

Under the assumption (\ref{9.110}), we can neglect the terms
containing $r^{-n}$  $(n\geq 1)$ in (\ref{9.102}). In addition, by
(\ref{9.110}) we have
\begin{eqnarray*}
-\sqrt{T}\left(\frac{\eta^2-\mu}{1-\eta^2}-\frac{1-\mu}{1-\eta^2}\frac{r^2_1}{r^2}\right)&=&
\sqrt{T}\left(1-\frac{1-\mu}{1-\eta^2}\frac{r^2-r^2_1}{r^2}\right)\\
&\simeq&\sqrt{T}(1-(1-\mu )(r-r_1)).
\end{eqnarray*}
Let
\begin{equation}
\alpha =\frac{\eta^2-\mu}{1-\eta^2}.\label{9.111}
\end{equation}
Replacing $u_{\theta}$ by $\sqrt{\alpha}u_{\theta}$, and assuming
the perturbations are axi-symmetric and are independent of $\theta$,
we obtain from (\ref{9.102}):
\begin{eqnarray}
\left.\begin{aligned} &\frac{\partial u_z}{\partial
t}+(\tilde{u}\cdot\nabla )u_z=\Delta u_z-\frac{\partial p}{\partial
z},\\
&\frac{\partial u_r}{\partial t}+(\tilde{u}\cdot\nabla )u_r=\Delta
u_r-\frac{\partial p}{\partial r}+\sqrt{\alpha T}(1-(1-\mu
)(r-r_1))u_{\theta},\\
&\frac{\partial u_{\theta}}{\partial t}+(\tilde{u}\cdot\nabla
)u_{\theta}=\Delta u_{\theta}+\sqrt{\alpha T}u_r,\\
&\frac{\partial u_r}{\partial r}+\frac{\partial u_z}{\partial z}=0,
\end{aligned}
\right.\label{9.112}
\end{eqnarray}
where
$$\Delta =\frac{\partial^2}{\partial r^2}+\frac{\partial^2}{\partial
z^2},\ \ \ \ (\tilde{u}\cdot\nabla )=u_r\frac{\partial}{\partial
r}+u_z\frac{\partial}{\partial z}.$$

In this case, the spatial domain is $M=(r_1,r_1+1)\times (0,L)$. For
convenience, we  consider here the Dirichlet boundary condition
\begin{equation}
u|_{\partial M}=0.\label{9.113}
\end{equation}
The initial value condition is axisymmetric, and given by
\begin{equation}
u=u_0(r,z)\ \ \ \ \text{at}\ \ \ \ t=0.\label{9.114}
\end{equation}

The linearized equations of (\ref{9.112}) read
\begin{eqnarray}
\left.\begin{aligned} &-\Delta u_z+\frac{\partial p}{\partial
z}=0,\\
&-\Delta u_r+\frac{\partial p}{\partial r}=\lambda
u_{\theta}-\lambda (1-\mu )(r-r_1)u_{\theta},\\
&-\Delta u_{\theta}=\lambda u_r,\\
&\frac{\partial u_r}{\partial r}+\frac{\partial u_z}{\partial z}=0,
\end{aligned}
\right.\label{9.115}
\end{eqnarray}
where $\lambda =\sqrt{\alpha T},T$ is the Taylor number given by
(\ref{9.97}).

Let $\lambda_1>0$ be the first eigenvalue of (\ref{9.115}) with
(\ref{9.113}). We call
\begin{equation}
T_c=\lambda^2_1/\alpha ,\label{9.116}
\end{equation}
the critical Taylor number, where $\alpha$  is given by (\ref{9.111}).

As $\mu\rightarrow 1$,   equations (\ref{9.115}) are reduced to the following symmetric
linear equations:
\begin{equation}
\left.\begin{aligned} &-\Delta u_z+\frac{\partial p}{\partial
z}=0,\\
&-\Delta u_r+\frac{\partial p}{\partial r}=\lambda u_{\theta},\\
&-\Delta u_{\theta}=\lambda u_r,\\
&\frac{\partial u_z}{\partial z}+\frac{\partial u_r}{\partial r}=0.
\end{aligned}
\right.\label{9.117}
\end{equation}

Let the first eigenvalue $\lambda_0>0$ of (\ref{9.117}) with
(\ref{9.113}) have multiplicity $m\geq 1$, the corresponding
eigenfunctions be $v_i$  $(i=1,\cdots,m)$, and the corresponding
eigenspace be
$$E_0=\text{span}\{v_i\ |\ 1\leq i\leq m\}.$$

We remark here that under conditions (\ref{9.99}) and (\ref{9.110}),
the condition $\mu\rightarrow 1$ can be equivalently replaced by
\begin{equation}
r_1=(2+\delta )/(1-\mu ),\label{9.118}
\end{equation}
for some $\delta >0$. In this case the parameter $\alpha$ in
(\ref{9.111}) is
$$\alpha =(\eta^2-\mu )/(1-\eta^2)\simeq\delta /2.$$

When the conditions (\ref{9.110}) and (\ref{9.118}) hold true,
$\mu\rightarrow 1$ and $r_1\rightarrow\infty$. In this case the
equations (\ref{9.145}) are replaced by (\ref{9.112}), and the
linearized equations of (\ref{9.112}) reduces to the symmetric
linear system (\ref{9.117}). For the approximate problem
(\ref{9.117}) with (\ref{9.146}), we use $R_0$ to denote the number
$R$ defined by (\ref{9.167}) and (\ref{9.168}):
$$R_0=\frac{1}{\|\psi_1\|^2}\left[(G(\Phi
,\psi_1),\psi_1)+G(\psi_1,\Phi ),\psi_1)_H\right].$$ Here $\psi_1$
is given by (\ref{9.158}) with $(h,\varphi )$ satisfying
$$\left.
\begin{aligned}
&(D^2-a^2)^2h=\lambda_0\varphi ,\\
&(D^2-a^2)\varphi =-\lambda_0h,\\
&h=Dh=0,\varphi =0 && \text{at}\ \ \ \ r=1,\eta ,
\end{aligned}
\right.
$$ 
and $\Phi$ is defined by
\begin{equation}
\left.
\begin{aligned} &(A-\lambda_0B_0)\Phi =G(\psi_1,\psi_1),\\
&B_0\Phi =P(0,\Phi_{\theta},\Phi_r).
\end{aligned}
\right.\label{9.188}
\end{equation}
By (\ref{9.150}) we have
\begin{eqnarray*}
&&(G(\Phi ,\psi_1),\psi_1)_H=0,\\
&&(G(\psi_1,\Phi ),\psi_1)_H=-(G(\psi_1,\psi_1),\Phi )_H.
\end{eqnarray*}
Hence, we infer from (\ref{9.188}) that
\begin{eqnarray*}
R_0&=&-\frac{1}{\|\psi_1\|^2}(G(\psi_1,\psi_1),\Phi )\\
&=&-\frac{1}{\|\psi_1\|^2}((A-\lambda_0B_0)\Phi ,\Phi ).
\end{eqnarray*}
We see that $A-\lambda_0B_0$ is symmetric and semi-positive
definite, and
$$
G(\psi_1,\psi_1)\bot \text{Ker}(A-\lambda_0B_0),\ \ \ \
\Phi\bot\text{Ker}(A-\lambda_0B_0).
$$ 
Therefore it follows that
$$
R_0=-\frac{1}{\|\psi_1\|^2}((A-\lambda_0B_0)^{1/2}\Phi
,(A-\lambda_0B_0)^{1/2}\Phi )_H<0.
$$ 
On the other hand, it
is known that the number $R(\mu )$ in (\ref{9.167}) is continuous on
$\mu$, and
$$R(\mu )\rightarrow R_0\ \ \ \ \text{as}\ \ \ \ \mu\rightarrow 1.$$
Hence we derive the following conclusion.

\bt\la{t9.12}
For the Taylor problem
(\ref{9.145})-(\ref{9.147}) there exist $\mu_0<1$ and $0<\eta_0<1$
such that for any $\mu_0<\mu <1$ and $\eta_0<\eta <1$ with $\mu
<\eta^2$, the parameter $R=R(\mu ,\eta )$ defined by (\ref{9.167})
is negative, i.e.,
$$
R(\mu ,\eta )<0,\ \ \ \ \forall\mu_0<\mu <1,\ \ \ \ \eta_0<\eta  <1.
$$ 
Consequently, the conclusions in Theorem \ref{t9.10} hold true.
\et

\section{Formation of the Taylor vortices  and Structural Stability}
Assertions (5) and (6) in Theorem \ref{t9.10} provide an asymptotic
structure of the solutions in the physical space when the gap
$r_2-r_1$ is small, as observed in the experiments. However, for
general parameters $\eta$ and $\mu$ we can not give the precisely
theoretic results, and only present some qualitative description.
Here we consider two general cases as follows.

{\sc Case: $\mu\geq 0$.}
 Following \cite{yudovich66,velte}, for the
eigenvector $(h,\varphi )$ of (\ref{9.160}) the function $h$ can be
taken as positive and has a unique maximum point in the interval
$(\eta ,1)$. Therefore, for the eigenvectors defined by
(\ref{9.158}) and (\ref{9.159}), the vector fields $(\psi_z,\psi_r)$
and $(\tilde{\psi}_z,\tilde{\psi}_r)$ are divergence-free and have
the topological structure as shown in Figure \ref{f9.13}. Hence to obtain
Assertions (5) and (6) in Theorem \ref{t9.10} for any $\mu\geq 0$ it
suffices to prove that
\begin{equation}
h^{\prime\prime}(r)\neq 0\ \ \ \ \text{at}\ \ \ \ r=r_0,1,\eta
,\label{9.189}
\end{equation}
where $r_0\in (\eta ,1)$ is the maximum point of $h$. We conjecture
that the property (\ref{9.189}) for the eigenvector $(h,\varphi )$
of (\ref{9.160}) is valid for all $0\leq\mu <1$ and $0<\eta <1$.

\medskip

{\sc Case: $\mu <0$.} In this case, the situation is different. Numerical
results show that the vector field $(\psi_z,\psi_r)$ in
(\ref{9.158}) has $k\geq 2$ vortices in the radial direction, called
the Taylor vortices, which has the topological structure as shown in
Figure \ref{f9.15}; see \cite{chandrasekhar}. 
This type of structure is
structurally unstable. However, as discussed in \cite{MW00a},
under a perturbation either in space $H$ or in
$$
\tilde{H}=\left\{(u_z,u_r,u_{\theta})\in H|\
\int^L_0\int^1_{\eta}ru_zdzdr=0\right\},
$$ 
there are only finite types of
stable structures. In particular, if the vector field
$(\psi_z,\psi_r)$ in (\ref{9.158}) is $D$-regular, i.e., $h(r)$
satisfies (\ref{9.189}), then there is only one class of stable
structures regardless of the orientation. For example, when $u_0\in
H/(\Gamma\cup\tilde{H})$, the asymptotic structure of the solution
$u(t,u_0)$ of (\ref{9.145})-(\ref{9.147}) is as shown in Figure
\ref{f9.16}, and when $u_0\in\tilde{H}  \setminus \Gamma$, the asymptotic structure of
the solution $u(t,u_0)$ is as shown in Figure \ref{f9.17}. It is clear that
the class of structures illustrated by Figure \ref{f9.16} is different from
that illustrated by Figure \ref{f9.17}. The first one has a cross the channel traveling
flow in  the $z$-direction and the second one does not have such a cross the channel flow.
\begin{SCfigure}[25][t]
  \centering
  \includegraphics[width=0.4\textwidth]{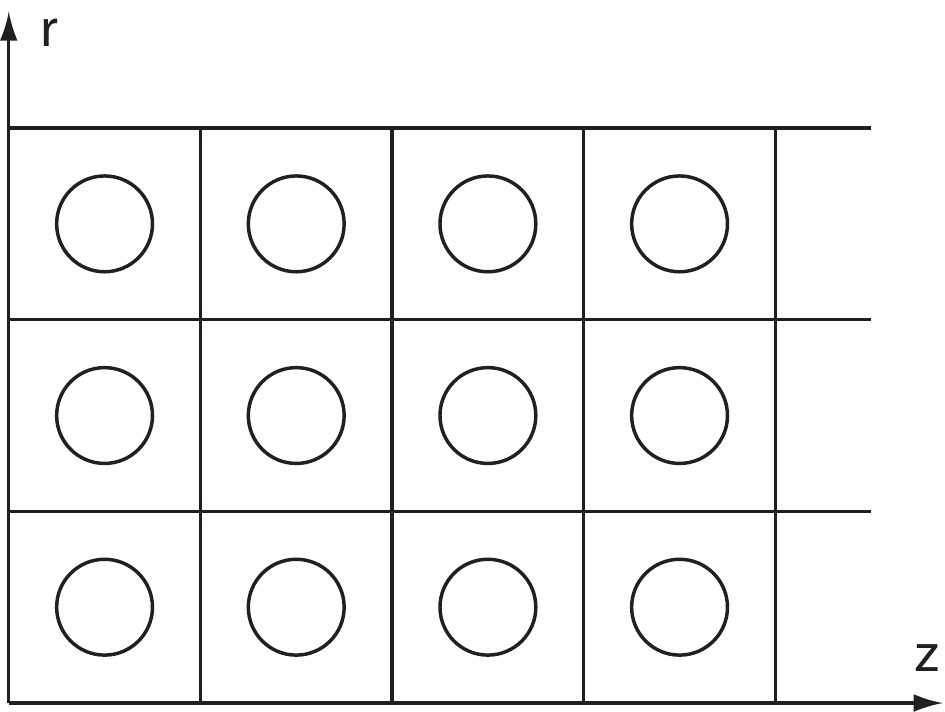}
  \caption{$(\psi_z,\psi_r)$ has $k$ vortices in
$r$-direction.}\la{f9.15}
 \end{SCfigure}

\begin{SCfigure}[25][t]
  \centering
  \includegraphics[width=0.4\textwidth]{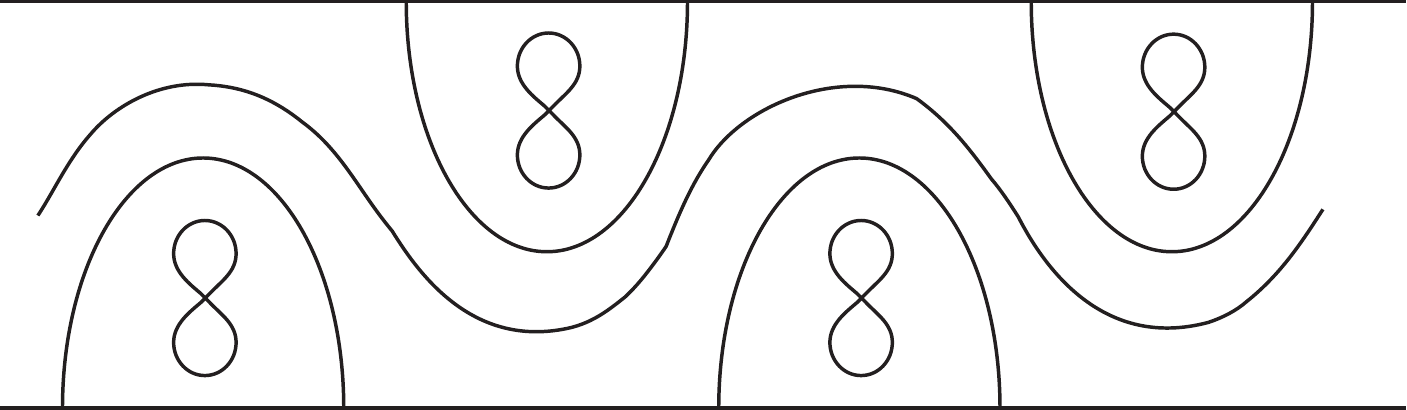}
  \caption{The stable structure with a perturbation in
space $H/(\Gamma\cup\tilde{H}).$}\la{f9.16}
 \end{SCfigure}
 
\begin{SCfigure}[25][t]
  \centering
  \includegraphics[width=0.4\textwidth]{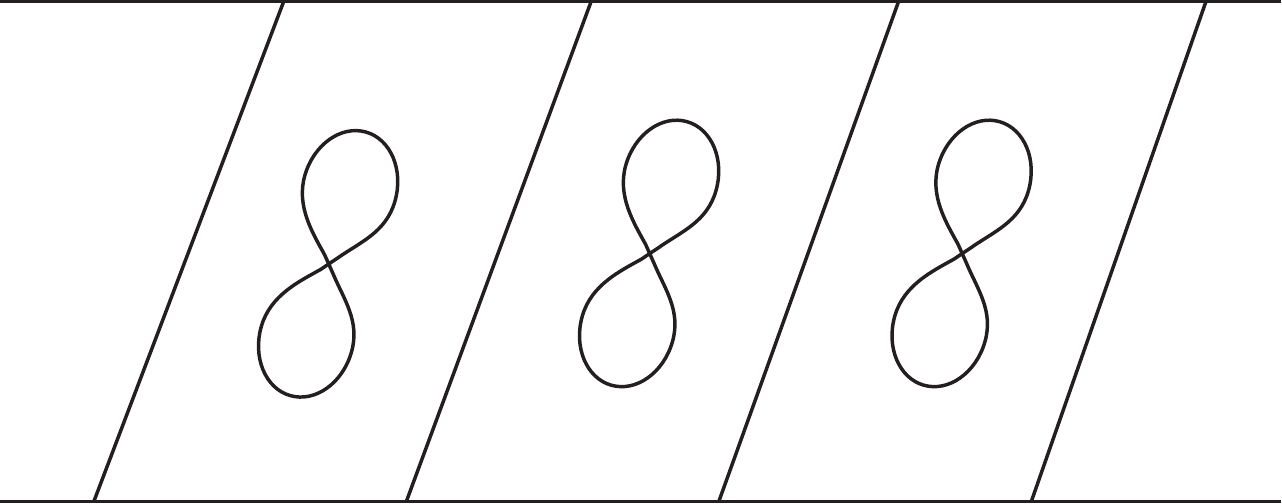}
  \caption{The stable structure with a perturbation in
space $\tilde{H}\setminus \Gamma$.}\la{f9.17}
 \end{SCfigure}

\section{Proof of Main Theorems}
\subsection{Proof of Theorem \ref{t9.10}}
 We shall prove this theorem in the  following several steps.

\medskip

{\sc  Step 1.}
 We claim that the problem (\ref{9.145})-(\ref{9.147})
bifurcates from $(u,\lambda )=(0,\lambda_0)$ to a circle $S^1$ which
consists of stead states.

It is easy to see that the problem (\ref{9.145}) with (\ref{9.146})
is invariant for the transition in the $z$-direction
$$u(z,r,t)\rightarrow u(z+z_0,r,t)\qquad \text{ for}  z_0\in \R^1.$$
Therefore, if $u_0$ is a steady state solution of (\ref{9.145}) with
(\ref{9.146}),  then for any $z_0\in \R^1$ the function $u_0(z+z_0,r)$ is also a steady state solution. We can see that the set
$$\Sigma =\{u_0(z+z_0,r)|\ z_0\in \R^1\}$$
is homeomorphic to a circle $S^1$ in $H_1$ for any $u_0\in H_1$.
Hence, the singular points of (\ref{9.145}) with (\ref{9.146})
appear as a circle.

It is known in \cite{yudovich66,velte} that there exist
singular points bifurcated from $(u,\lambda )=(0,\lambda_0)$. Thus
this claim is proved.

\medskip

{\sc  Step 2.  Reduction to the center manifold.}
 We shall use the construction of center manifold functions  to derive
the reduced equations of (\ref{9.149}) given by
\begin{equation}
\left.\begin{aligned} 
&\frac{dx}{dt}=\beta_1(\lambda)x
+\frac{(G(u),\psi^*_{1\lambda})_H}{ (\psi_{1\lambda},\psi^*_{1\lambda})_H},\\
&\frac{dy}{dt}=\beta_1(\lambda)y
+\frac{(G(u),\tilde{\psi}^*_{1\lambda})_H}{ (\tilde{\psi}_{1\lambda},\tilde{\psi}^*_{1\lambda})_H},
\end{aligned}
\right.\label{9.169}
\end{equation}
where $\psi_{1\lambda}$ and $\tilde{\psi}_{1\lambda}$ are the
eigenvectors of (\ref{9.153}) corresponding to $\beta_1(\lambda )$
near $\lambda =\lambda_0$ with
\begin{equation}
\left.\begin{aligned}
&\lim_{\lambda\rightarrow\lambda_0}\psi_{1\lambda}=\psi_1 &&
(\psi_1\ \text{as\ in}\ (\ref{9.158})),\\
&\lim_{\lambda\rightarrow\lambda_0}\tilde{\lambda}_{1\lambda}=\tilde{\psi}_1&&
(\tilde{\psi}_1\ \text{as\ in}\ (\ref{9.159})),
\end{aligned}
\right.\label{9.170}
\end{equation}
and $\psi^*_{1\lambda}$ and $\tilde{\psi}^*_{1\lambda}$ are the dual
eigenvectors of $\psi_{1\lambda}$ and $\tilde{\psi}_{1\lambda}$
satisfying
\begin{equation}
\left.\begin{aligned}
&\lim\limits_{\lambda\rightarrow\lambda_0}\psi^*_{1\lambda}=\psi^*_1 &&
(\psi^*_1\ \text{as\ in}\ (\ref{9.161})),\\
&\lim\limits_{\lambda\rightarrow\lambda_0}\tilde{\psi}^*_{1\lambda}=\tilde{\psi}^*_1&&
(\tilde{\psi}_1\ \text{as\ in}\ (\ref{9.162})).
\end{aligned}
\right.\label{9.171}
\end{equation}

Let $\Psi :E_0\rightarrow E^{\bot}_0$ be the center manifold
function of (\ref{9.149}) at $\lambda =\lambda_0$, where
\begin{eqnarray*}
&&E_0=\text{span}\{\psi_1,\tilde{\psi}_1\},\\
&&E^{\bot}_0=\{u\in H|\ (u,\psi^*_1)_H=0,(u,\tilde{\psi}^*_1)_H=0\}.
\end{eqnarray*}
Let $u_0=x\psi_1+y\tilde{\psi}_1\in E_0$. Then it is easy to check
that $G(u_0)\in E^{\bot}_0$. Hence, by the center manifold approximation formula in \cite{MW09c, ptd},  we find that
\begin{equation}
\left.\begin{aligned} &\Psi =\phi
(x,y)+o(|x|^2+|y|^2)+O(\beta_1(\lambda )(|x|^2+|y|^2)),\\
&-L_{\lambda}\phi
=G(\psi_1,\psi_1)x^2+G(\tilde{\psi}_1,\tilde{\psi}_1)y^2+(G(\psi_1,\tilde{\psi}_1)+G(\tilde{\psi}_1,
\psi_1))xy.
\end{aligned}
\right.\label{9.172}
\end{equation}
On the center manifold, $u=u_0+\Phi (u_0)$. Therefore from
(\ref{9.169})-(\ref{9.172}) we obtain the reduced equations of
(\ref{9.149}) to the center manifold as follows:
\begin{equation}
\left.
\begin{aligned}
\frac{dx}{dt} 
  =&\beta_1(\lambda )x+\frac{x}{\rho}(G(\phi
,\psi_1)+G(\psi_1,\phi ),\psi^*_1)_H  \\
&+\frac{y}{\rho}(G(\phi ,\tilde{\psi}_1)+G(\tilde{\psi}_1,\phi
),\psi^*_1)_H     \\
&+o(|x|^3+|y|^3)+\varepsilon_1(\lambda
)O(|x|^3+|y|^3),
\\
\frac{dy}{dt}=&\beta_1(\lambda )y+\frac{x}{\rho}(G(\phi,\psi_1)
+G(\psi_1,\phi ),\tilde{\psi}^*_1)_H\\
&+\frac{y}{\rho}(G(\phi ,\tilde{\psi}_1)+G(\tilde{\psi}_1,\phi
),\tilde{\psi}^*_1)_H\\
&+o(|x|^3+|y|^3)+\varepsilon_2(\lambda )O(|x|^3+|y|^3),
\end{aligned}\right.
\label{9.173}
\end{equation}
where $\rho
=(\psi_1,\psi^*_1)_H=(\tilde{\psi}_1,\tilde{\psi}^*_1)_H$, and
$$\lim_{\lambda\to \lambda_0} \varepsilon_i(\lambda )= 0\qquad  (i=1,2).$$

Furthermore, direct calculation shows that
\begin{align*}
&G(\psi_1,\psi_1)=P\psi_0+P\psi_2, &&G(\tilde{\psi}_1,\tilde{\psi}_1)=P\psi_0-P\psi_2,\\
&G(\psi_1,\tilde{\psi}_1)=P\tilde{\psi}_0+P\tilde{\psi}_2,
&&G(\tilde{\psi}_1,\psi_1)=-P\tilde{\psi}_0+P\tilde{\psi}_2.
\end{align*}
where $P:L^2(M)^3 \rightarrow H$ is the Leray projection, and
\begin{align*}
&
\psi_0=-\left\{\begin{aligned}
&0,\\
&\frac12(a^2hD_*h+a^2hDh-\frac{1}{r}\varphi^2),\\
&\frac{a}2(\varphi D_*h+hD\varphi +\frac{1}{r}h\varphi ),
\end{aligned}
\right.\\
&
\psi_2=-\left\{\begin{aligned} &\frac{a}{2}\sin
2az((D_*h)^2-hDD_*h),\\
&\frac{1}{2}\cos 2az(a^2hDh-a^2hD_*h-\frac{1}{r}\varphi^2),\\
&\frac{a}{2}\cos 2az(hD\varphi -\varphi D_*h+\frac{1}{r}\varphi h),
\end{aligned}
\right.\\
&
\tilde{\psi}_0=-\left\{\begin{aligned}
&\frac{a}{2}((D_*h)^2+hDD_*h),\\
&0,\\
&0,
\end{aligned}
\right.\\
&
\tilde{\psi}_2=-\left\{\begin{aligned} &-\frac{a}{2}\cos
2az((D_*h)^2-hDD_*h),\\
&\frac12\sin 2az(a^2hDh-a^2hD_*h-\frac{1}{r}\varphi^2),\\
&\frac{a}{2}\sin 2az(hD\varphi -\varphi D_*h+\frac{1}{r}\varphi h).
\end{aligned}
\right.
\end{align*}
Thus, (\ref{9.172}) is rewritten as
\begin{equation}
(A-\lambda_0B)\phi
=(x^2+y^2)P\psi_0+(x^2-y^2)P\psi_2+2xy P\tilde{\psi}_2.\label{9.174}
\end{equation}

Let
 \begin{align}
& 
\phi =-[(x^2+y^2)\phi_0+(x^2-y^2)\phi_2+2xy\tilde{\phi}_2], \label{9.175}\\
&
 \left\{
 \begin{aligned}
&\phi_0=(0,0,\varphi_0),\\
&\phi_2=(-1/2\sin 2az\varphi_z,\cos 2az\varphi_r,\cos
2az\varphi_{\theta}),\\
&\tilde{\phi}_2=(1/2\cos 2az\tilde{\phi}_z,\sin
2az\tilde{\phi}_r,\sin 2az\tilde{\phi}_{\theta}).
\end{aligned}
\right.\label{9.176}
\end{align}
Then we deduce from (\ref{9.174}) and (\ref{9.175}) that
$$
\varphi_z=\tilde{\varphi}_z,\ \ \ \ \varphi_r=\tilde{\varphi}_r,\
\ \ \ \varphi_{\theta}=\tilde{\varphi}_{\theta},
$$ and
$(\varphi_z,\varphi_r,\varphi_{\theta})$ satisfies
\begin{eqnarray*}
&&(DD_*-4a^2)\varphi_r+4a^2\lambda_0(\frac{1}{r^2}-\kappa
)\varphi_{\theta}=4a^2H_2+2aDH_1,\\
&&(DD_*-4a^2)\varphi_{\theta}+\lambda_0\kappa\varphi_r=H_3,\\
&&\varphi_z=\frac{1}{2a}D_*\varphi_r,\\
&&\varphi_z=0,\varphi_{\theta}=0,\varphi_r=D\varphi_r=0\ \ \ \
\text{at}\ \ \ \ r=\eta ,1,
\end{eqnarray*}
where $H_1,H_2$, and $H_3$ are given by
\begin{equation}
\left.
\begin{aligned} &H_1=a((D_*h)^2-hDD_*h),\\
&H_2=a^2hDh-a^2hD_*h-\frac{1}{r}\varphi^2,\\
&H_3=a(hD\varphi -\varphi D_*h+\frac{1}{r}\varphi h).
\end{aligned}
\right.\label{9.177}
\end{equation}
Based on (\ref{9.176}) we find
\begin{eqnarray*}
&&(G(\tilde{\psi}_1,\phi_i)+G(\phi_i,\tilde{\psi}_1),\psi^*_1)_H=0\
\ \ \ \text{for}\ \ \ \ i=0,2,\\
&&(G(\psi_1,\phi_i)+G(\phi_i,\psi_1),\tilde{\psi}^*_1)_H=0\ \ \ \
\text{for}\ \ \ \ i=0,2,\\
&&(G(\tilde{\psi}_1,\tilde{\phi}_2)+G(\tilde{\phi}_2,\tilde{\psi}_1),\tilde{\psi}^*_1)_H=0,\\
&&(G(\psi_1,\tilde{\phi}_2)+G(\tilde{\phi}_2,\psi_1),\psi^*_1)_H=0.
\end{eqnarray*}
Then, putting (\ref{9.175}) into (\ref{9.173}), we deduce that 
\begin{equation}
\left.
\begin{aligned}
\frac{dx}{dt}= \beta_1x
&
-\frac{1}{\rho}x(x^2+y^2)(G(\phi_0,\psi_1)+
G(\psi_1,\phi_0),\psi^*_1)_H  \\
&-\frac{1}{\rho}x(x^2-y^2)(G(\phi_2,\psi_1)+G(\psi_1,\phi_2),\psi^*_1)_H \\
&-\frac{2}{\rho}xy^2(G(\tilde{\phi}_2,\tilde{\psi}_1)+G(\tilde{\psi}_1,\tilde{\phi}_2),\psi^*_1)_H   \\
&+ o(|x|^3+|y|^3)+\varepsilon_1(\lambda)O(|x|^3+|y|^3),
\\
\frac{dy}{dt}=\beta_1 y
& -\frac{1}{\rho}y(x^2+y^2)(G(\phi_0,\tilde{\psi}_1)+G(\tilde{\psi}_1,\phi_0),
\tilde{\psi}^*_1)_H\\
&-\frac{1}{\rho}y(x^2-y^2)(G(\phi_2,\tilde{\psi}_1)+G(\tilde{\psi}_1,\phi_2),\tilde{\psi}^*_1)_h\\
&-\frac{2}{\rho}yx^2(G(\tilde{\phi}_2,\psi_1)+G(\psi_1,\tilde{\phi}_2),\tilde{\psi}^*_1)_H\\
&+o(|x|^2+|y|^2)+\varepsilon_2(\lambda )O(|x|^3+|y|^3).
\end{aligned}\right.\label{9.178}
\end{equation}

Direct computation yields
\begin{eqnarray*}
(G(\phi_2,\psi_1)+G(\psi_1,\phi_2),\psi^*_1)_H&=&(G(\tilde{\phi}_2,\tilde{\psi}_1)+G(\tilde{\psi}_1,
\tilde{\phi}_2),\psi^*_1)_H\\
&=&(G(\phi_2,\tilde{\psi}_1)+G(\tilde{\psi}_1,\phi_2),\tilde{\psi}^*_1)_H\\
&=&(G(\tilde{\phi}_2,\psi_1)+G(\psi_1,\tilde{\phi}_2),\tilde{\psi}^*_1)_H.
\end{eqnarray*}
Hence (\ref{9.178}) can be  rewritten as
\begin{equation}
\left.
\begin{aligned}
&\frac{dx}{dt}=\beta_1x+Rx(x^2+y^2)+o(|x|^3+|y|^3)+\varepsilon_1(\lambda
)O(|x|^3+|y|^3),\\
&\frac{dy}{dt}=\beta_1y+Ry(x^2+y^2)+o(|x|^3+|y|^3)+\varepsilon_2(\lambda
)O(|x|^3+|y|^3),
\end{aligned}
\right.\label{9.179}
\end{equation}
where
\begin{equation}
R=-\frac{1}{\rho}(G(\phi_0+\phi_2,\psi_1)+G(\psi_1,\phi_2+\phi_2),\psi^*_1)_H.\label{9.180}
\end{equation}

On the other hand,  we infer  from (\ref{9.174}) and (\ref{9.176}) that
$$(A-\lambda_0B)\phi_i=P\psi_i \qquad \text{ for }  i=0,2,$$
Hence we find
$$\left.
\begin{aligned}
&\Phi =-(\phi_0+\phi_2),\\
&(A-\lambda_0B)\Phi =G(\psi_1,\psi_1)=P\psi_0+P\psi_2.
\end{aligned}
\right.
$$ 
Thus  the number (\ref{9.180}) is the same as that in (\ref{9.167}).

\medskip

{\sc Step 3. Proof of Assertions (1)-(4).} 
When $R<0$, $(x,y)=0$ is locally
asymptotically stable for (\ref{9.179}) at $\lambda =\lambda_0$.
Therefore $u=0$ is a locally asymptotically stable singular point of
(\ref{9.149}). By the attractor bifurcation theorem, Theorem~6.1 in p. 153 in \cite{b-book}, the problem
(\ref{9.145})-(\ref{9.147}) bifurcates from $(u,\lambda
)=(0,\lambda_0)$ to an attractor ${\mathcal{A}}_{\lambda}$ which
attracts an open set $U\setminus \Gamma$, and Assertions (1), (3) and (4)
hold true.

In addition, the nonlinear terms in (\ref{9.179}) satisfy the
coercive condition in the $S^1$-attractor bifurcation theorem
(Theorem~5.10 in \cite{b-book}) and the conclusion in Step 1., and Assertion
(2) follows.

\medskip

{\sc Step 4. Attraction in $C^r$-norm.}
 It is known that for any
initial value $u_0\in H$ there is a time $t_0>0$ such that the
solution $u(t,u_0)$ of (\ref{9.145})-(\ref{9.147}) is analytic for
$t>t_0$, and uniformly bounded in $C^r$-norm for any $r\geq 1$; see
Theorem 1 in \cite{MW06b}. Hence, by Assertion (4), for any $u_0\in U\setminus \Gamma$
we have 
\begin{equation}
\lim\limits_{t\rightarrow\infty}\min\limits_{v_0\in{\mathcal{A}}_{\lambda}}
\|u(t,u_0)-v_0\|_{C^r}=0.\label{9.181}
\end{equation}

\medskip

{\sc  Step 5. Structure of solutions in ${\mathcal{A}}_{\lambda}$}. By
Assertion (3), for any steady state solution
$u_0=(u_z,u_r,u_{\theta})\in{\mathcal{A}}_{\lambda}$, the vector
field $\tilde{u}=(u_z,u_r)$ of $u_0$ can be expressed as
\begin{equation}
\left.
\begin{aligned} 
&u_z=\gamma\cos
a(z+z_0)D_*h(r)+w_1(z,r,\beta_1),\\
&u_r=a\gamma\sin a(z+z_0)h(r)+w_2(z,r,\beta_1),
\end{aligned}
\right.\label{9.182}
\end{equation}
for some $z_0\in \R^1$, where 
$$
\gamma =|\beta_1(\lambda )/R|^{1/2},\qquad  w_i=o(|\beta_1|^{1/2}) 
\qquad \text{ for }  i=1,2.
$$

As in the proof of Theorem 4.1 in \cite{MW07a}, we  deduce that the
vector field (\ref{9.182}) is $D$-regular for all $0<\lambda
-\lambda_0<\varepsilon$ for some $\varepsilon >0$. Moreover, the
first order vector field in (\ref{9.182})
\begin{equation}
(v_z,v_r)=(\gamma\cos a(z+z_0)D_*h(r),a\gamma\sin
a(z+z_0)h(r))\label{9.183}
\end{equation}
has the topological structure as shown in Figure \ref{f9.13}.

Furthermore, it is easy to check that the space
$$\tilde{H}=\left\{u=(u_z,u_r,u_{\theta})\in H|\ \int_Mru_zdrdz=0\right\}$$
is invariant for the operator $L_{\lambda}+G$ defined by
(\ref{9.148}). To see this, since  $u$ is $z$-periodic and $u=0$ at
$r=1,\eta$, we have
\begin{eqnarray*}
\int^L_0\int^1_{\eta}r(\tilde{u}\cdot\nabla
)u_zdrdz&=&\int^L_0\int^1_{\eta}ru_r\frac{\partial u_z}{\partial
r}drdz\\
&=&-\int^L_0\int^1_{\eta}\frac{\partial (ru_r)}{\partial r}u_zdrdz\\
&=&\int^L_0\int^1_{\eta}r\frac{\partial u_z}{\partial z}u_zdrdz\\
&=&0\qquad  \forall u\in H,
\end{eqnarray*}
$$\int^L_0\int^1_{\eta}r\Delta u_zdrdz=0,\ \ \ \ \forall
u\in\tilde{H}.
$$ 
Thus we see that $\tilde{H}$ is invariant for
$L_{\lambda}+G$.

Therefore, for the vector field (\ref{9.181}) we have
$$\int_Mru_zdrdz=0.$$
By The Connection Lemma and the orbit-breaking method in \cite{amsbook}, it implies that the vector field (\ref{9.182}) is
topologically equivalent to its first order field (\ref{9.183}) for
$0<\lambda -\lambda_0<\varepsilon$.

\medskip

{\sc  Step 6. Proof of Assertions (5) and (6).}
For any initial value
$u_0\in U\setminus (\Gamma\cup\tilde{H})$, we have
\begin{equation}\label{9.184}
u_0=\sum_{k=1}^\infty\alpha_ke_k+w_0,
\end{equation}
where $ w_0\in\tilde{H}$, and for any $k=1, 2, \cdots, $
 \begin{align*}
& e_k=(\tilde{e}_k(r),0,0) ,\\
&  \int^1_{\eta}\tilde{e}_k(r)dr\neq 0 ,
\end{align*}
and  $\tilde{e}_k(r)$ satisfies  that
$$\left.
\begin{aligned}
&D_*D\tilde{e}_k=-\rho_k\tilde{e}_k,\\
&\tilde{e}_k|_{r=\eta, 1}=0,\\
& 0<\rho_1<\rho_2<\cdots.
\end{aligned}
\right.$$ 
Make the decomposition
\begin{eqnarray*}
&&H_1=E\bigoplus\tilde{H}_1\ \ \ \ (\tilde{H}_1=H_1\cap\tilde{H}),\\
&&H=E\bigoplus\tilde{H},\\
&&E=\text{span}\{e_1,e_2,\cdots\}.
\end{eqnarray*}
Then the equation (\ref{9.149}) can be decomposed into
\begin{equation}
\left.
\begin{aligned} 
&\frac{de}{dt}=L_{\lambda}e &&  e\in E,\\
&\frac{dw}{dt}=L_{\lambda}w+G(w)&& w\in\tilde{H}_1,\\
&(e(0), w(0))=(\sum_k\alpha_ke_k, w_0).
\end{aligned}
\right.\label{9.185}
\end{equation}
It is obvious that
$$L_{\lambda}e_k=-\rho_ke_k.$$
Hence for the initial value (\ref{9.184}), the solution $u(t,u_0)$
of (\ref{9.185}) can be expressed as
\begin{equation}
\left.
\begin{aligned}
&u(t,u_0)=\sum\limits_k\alpha_ke^{-\rho_kt}e_k+w(t,u_0),\\
&\int^1_{\eta}e_k(r)dr\neq 0.
\end{aligned}
\right.\label{9.186}
\end{equation}
By (\ref{9.181}) we have
$$\lim_{t\to\infty} \|w(t,u_0)-v_0\|_{C^r} = 0,
\qquad  v_0\in{\mathcal{A}}_{\lambda},
$$ which
implies by Step 5 that $w(t,u_0)$ is topologically equivalent to
(\ref{9.183}) for $t>0$ sufficiently large, i.e., $w(t,u_0)$ has the
topological structure as shown in Figure \ref{f9.13}.

By the structural stability theorem, Theorem~2.2.9 and Lemmas 2.3.1 and 2.3.3 (connection lemmas) in \cite{amsbook},  we  infer from
(\ref{9.186}) that the vector field in (\ref{9.186}) is
topologically equivalent to either the structure as shown in Figure
\ref{f9.12}(a) or the structure as shown in \ref{f9.12}(b), dictated by  the sign of
$\alpha_{k_0}$ in (\ref{9.184}) with $k_0=\min\{k|\alpha_k\neq 0\}$.
Thus Assertion (5) is proved.

Assertion (6) can be derived by the invariance of $\tilde{H}$ under
the operator $L_{\lambda}+G$ and the structural stability theorem
with perturbation in $\tilde{H}$, in the same fashion as in the proof of 
Theorem~2.2.9 in \cite{amsbook} by using the Connection Lemma.

The proof of Theorem \ref{t9.10} is complete.

\subsection{Proof of Theorem \ref{t9.11}}  
When $R>0$, by Theorem~A.2 in \cite{MW08c},  we infer
from the reduced equation (\ref{9.179}) that the transition of
(\ref{9.145})-(\ref{9.147}) is of Type-II. In the following,
we shall use the saddle-node bifurcation theorem, Theorem~A.7 in \cite{MW08c},  to
prove this theorem. Let
\begin{eqnarray*}
&&H^*=\{(u_z,u_r,u_{\theta})\in H|\ u_z(-z,r)=-u_z(z,r)\},\\
&&H^*_1=H_1\cap H^*.
\end{eqnarray*}

It is easy to see that the space $H^*$ is invariant under the action
of the operator $L_{\lambda}+G$ defined by (\ref{9.148}):
\begin{equation}
L_{\lambda}+G:H^*_1\rightarrow H^*,\label{9.187}
\end{equation}
and the first eigenvalue $\beta_1(\lambda )$ of
$L_{\lambda}:H^*_1\rightarrow H^*$ at $\lambda =\lambda_0(T=T_c)$ is
simple, with the first eigenvector $\psi_1$ given by (\ref{9.158}).
Hence, the number $R$ in (\ref{9.180}) is valid for the mapping
(\ref{9.186}), i.e., 
$$(G(x\psi_1+\Phi (x),\psi^*_1)=Rx^3+o(|x|^3).$$
Thus, it is readily to check that all conditions in  Theorem~A.7 in \cite{MW08c} are
fulfilled by the operator (\ref{9.187}). By Step 1 in the proof of
Theorem \ref{t9.10}, each singular point of (\ref{9.187}) generates a
singularity circle for $L_{\lambda}+G$ in $H$. Therefore, Theorem
\ref{t9.11} follows from  Theorem~A.7 in \cite{MW08c}.

The proof of Theorem \ref{t9.11} is complete.

\bibliographystyle{siam}

\begin{thebibliography}{10}

\bibitem{chandrasekhar}
{\sc S.~Chandrasekhar}, {\em Hydrodynamic and Hydromagnetic Stability}, Dover
  Publications, Inc., 1981.

\bibitem{dr}
{\sc P.~Drazin and W.~Reid}, {\em Hydrodynamic Stability}, Cambridge University
  Press, 1981.

\bibitem{kirch}
{\sc K.~Kirchg{\"a}ssner}, {\em Bifurcation in nonlinear hydrodynamic
  stability}, SIAM Rev., 17 (1975), pp.~652--683.

\bibitem{MW00a}
{\sc T.~Ma and S.~Wang}, {\em Structural evolution of the {T}aylor vortices},
  M2AN Math. Model. Numer. Anal., 34 (2000), pp.~419--437.
\newblock Special issue for R. Temam's 60th birthday.

\bibitem{b-book}
\leavevmode\vrule height 2pt depth -1.6pt width 23pt, {\em Bifurcation theory
  and applications}, vol.~53 of World Scientific Series on Nonlinear Science.
  Series A: Monographs and Treatises, World Scientific Publishing Co. Pte.
  Ltd., Hackensack, NJ, 2005.

\bibitem{amsbook}
\leavevmode\vrule height 2pt depth -1.6pt width 23pt, {\em Geometric theory of
  incompressible flows with applications to fluid dynamics}, vol.~119 of
  Mathematical Surveys and Monographs, American Mathematical Society,
  Providence, RI, 2005.

\bibitem{MW06b}
\leavevmode\vrule height 2pt depth -1.6pt width 23pt, {\em Stability and
  bifurcation of the {T}aylor problem}, Arch. Ration. Mech. Anal., 181 (2006),
  pp.~149--176.

\bibitem{MW07a}
\leavevmode\vrule height 2pt depth -1.6pt width 23pt, {\em Rayleigh-{B}\'enard
  convection: dynamics and structure in the physical space}, Commun. Math.
  Sci., 5 (2007), pp.~553--574.

\bibitem{MW08k}
\leavevmode\vrule height 2pt depth -1.6pt width 23pt, {\em Exchange of
  stabilities and dynamic transitions}, Georgian Mathematics Journal, 15:3
  (2008), pp.~581--590.

\bibitem{MW08c}
\leavevmode\vrule height 2pt depth -1.6pt width 23pt, {\em Cahn-hilliard
  equations and phase transition dynamics for binary systems}, Dist. Cont. Dyn.
  Systs., Ser. B, 11:3 (2009), pp.~741--784.

\bibitem{ptd}
\leavevmode\vrule height 2pt depth -1.6pt width 23pt, {\em Phase Transition
  Dynamics in Nonlinear Sciences}, submitted, 2009.

\bibitem{MW09c}
\leavevmode\vrule height 2pt depth -1.6pt width 23pt, {\em Dynamic transition
  theory for thermohaline circulation}, Physica D, 239:3-4 (2010),
  pp.~167--189.

\bibitem{taylor}
{\sc G.~I. Taylor}, {\em Stability of a viscous liquid contained between two
  rotating cylinders}, Philos. Trans. Royl London Ser. A, 223 (1923),
  pp.~289--243.

\bibitem{temam84}
{\sc R.~Temam}, {\em Navier-Stokes Equations, Theory and Numerical Analysis,
  3rd, rev. ed.}, North Holland, Amsterdam, 1984.

\bibitem{velte}
{\sc W.~Velte}, {\em Stabilit\"at and verzweigung station\"arer l\"osungen der
  davier-stokeschen gleichungen}, Arch. Rat. Mech. Anal., 22 (1966), pp.~1--14.

\bibitem{yudovich66}
{\sc V.~I. Yudovich}, {\em Secondary flows and fluid instability between
  rotating cylinders}, .\ Appl.\ Math.\ Mech., 30 (1966), pp.~822--833.

\end{thebibliography}

\end{document}